\documentclass[nonacm, sigconf, twocolumn]{acmart}

\renewcommand\footnotetextcopyrightpermission[1]{}
\setcopyright{none}

\usepackage{graphicx, subfig}
\usepackage[inline]{enumitem}
\usepackage{multirow}
\usepackage{xspace}

\usepackage{caption}
\usepackage[inline]{enumitem}

\usepackage{url}

\usepackage{breakurl}

\usepackage{amssymb}

\usepackage{xcolor}
\usepackage{tabularx,colortbl}
\usepackage{tikz}

\usepackage{hyperref}
\hypersetup{
  colorlinks=true,      
  linkcolor=blue,       
  citecolor=magenta,    
  filecolor=cyan,       
  urlcolor=red          
}

\usepackage[compact,small]{titlesec}
\usepackage[textfont={it}, font={small}, skip=1pt]{caption}

\usepackage{algorithm, algorithmicx,algpseudocode}
\algrenewcommand\algorithmicindent{0.5em}%

\newenvironment{denseitemize}{
\begin{itemize}[topsep=2pt, partopsep=0pt, leftmargin=1.5em]
  \setlength{\itemsep}{2pt}
  \setlength{\parskip}{0pt}
  \setlength{\parsep}{0pt}
}{\end{itemize}}

\def\ie{{i.e.\xspace}}
\def\eg{{e.g.\xspace}}
\def\etal{{et al.\xspace}}

\def\etc{etc.\xspace}

\def\meta{Meta\xspace}
\def\cxlsystem{CXL-System\xspace}
\def\cxlmem{CXL-Memory\xspace}
\def\Local{Local\xspace}
\def\local{local\xspace}
\def\remote{CXL-node\xspace}
\def\pebstool{Chameleon\xspace}
\def\TPP{TPP\xspace}
\def\collector{Collector\xspace}
\def\processor{Worker\xspace}

\newcommand{\rev}[1]{\textcolor{black}{#1}}

\newcommand*\circled[1]{\tikz[baseline=(char.base)]{
            \node[shape=circle,fill=black,draw,inner sep=1pt] (char) {\color{white}\bfseries#1};}}

\begin{document}
\sloppy
\date{}

\title[\TPP: Transparent Page Placement for CXL-Enabled Tiered-Memory]{\TPP: Transparent Page Placement for CXL-Enabled Tiered-Memory}

\author{Hasan Al Maruf}
\affiliation{
	\institution{University of Michigan}
 	\country{USA}
}
\author{Hao Wang}
\affiliation{
	\institution{NVIDIA}
 	\country{USA}
}

\author{Abhishek Dhanotia}
\affiliation{
	\institution{Meta Inc.}
 	\country{USA}
}

\author{Johannes Weiner}
\affiliation{
	\institution{Meta Inc.}
 	\country{USA}
}

\author{Niket Agarwal}
\affiliation{
	\institution{NVIDIA}
 	\country{USA}
}

\author{Pallab Bhattacharya}
\affiliation{
	\institution{NVIDIA}
 	\country{USA}
}

\author{Chris Petersen}
\affiliation{
	\institution{Meta Inc.}
 	\country{USA}
}

\author{Mosharaf Chowdhury}
\affiliation{
	\institution{University of Michigan}
 	\country{USA}
}

\author{Shobhit Kanaujia}
\affiliation{
	\institution{Meta Inc.}
 	\country{USA}
}

\author{Prakash Chauhan}
\affiliation{
	\institution{Meta Inc.}
 	\country{USA}
}

\renewcommand{\shortauthors}{Hasan Al Maruf \etal}

\begin{abstract}
The increasing demand for memory in hyperscale applications has led to memory becoming a large portion of the overall datacenter spend.  
The emergence of coherent interfaces like CXL enables main memory expansion and offers an efficient solution to this problem.
In such systems, the main memory can constitute different memory technologies with varied characteristics. 
In this paper, we characterize memory usage patterns of a wide range of datacenter applications across the server fleet of \meta.
We, therefore, demonstrate the opportunities to offload colder pages to slower memory tiers for these applications.
Without efficient memory management, however, such systems can significantly degrade performance. 

We propose a novel OS-level application-transparent page placement mechanism (\TPP) for CXL-enabled memory. 
\TPP employs a lightweight mechanism to identify and place hot/cold pages to appropriate memory tiers.
It enables a proactive page demotion from local memory to \cxlmem. This technique ensures a memory headroom for new page allocations that are often related to request processing and tend to be short-lived and hot.
At the same time, \TPP can promptly promote performance-critical hot pages trapped in the slow \cxlmem to the fast local memory, while minimizing both sampling overhead and unnecessary 
migrations.
\rev{\TPP works transparently without any application-specific knowledge and  can be deployed globally as a kernel release.}

We evaluate \TPP with diverse memory-sensitive workloads in the production server fleet with early samples of new x86 CPUs with CXL 1.1 support.
\rev{\TPP makes a tiered memory system performant as an ideal baseline (<1\% gap) that has all the memory in the local tier.}
It is 18\% better than today's Linux, and 5--17\% better than existing solutions including NUMA Balancing and AutoTiering.
\rev{Most of the \TPP patches have been merged in the Linux v5.18 release while the remaining ones are just pending for more discussion.}
\end{abstract}

\maketitle
\thispagestyle{empty}

\section{Introduction}
The surge in memory needs for datacenter applications~\cite{casefor, memorycost}, combined with the increasing DRAM cost and technology scaling challenges~\cite{tech-scaling, moor} has led to memory becoming a significant infrastructure expense in hyperscale datacenters. 
Non-DRAM memory technologies provide an opportunity to alleviate this problem by building tiered memory subsystems and adding higher memory capacity at a cheaper \$/GB point~\cite{fb-dcpmm, MyNVM, data-tiering, baidu-optane, azure-optane}. These technologies, however, have much higher latency vs. main memory and can significantly degrade performance when data is inefficiently placed in different levels of the memory hierarchy. Additionally, prior knowledge of application behavior and careful application tuning is required to effectively use these technologies. 
This can be prohibitively resource-intensive in hyperscale environments with varieties of rapidly evolving applications.

\begin{figure}[!t]
	\centering
	\subfloat[][\textbf{Without CXL}]{
	\label{fig:cxl-less-architecture}
		\includegraphics[width=0.45\columnwidth]{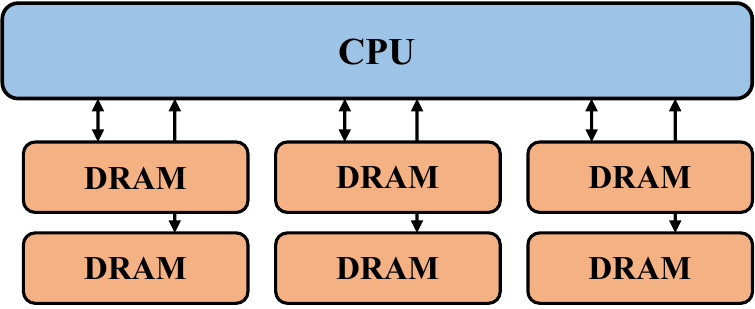}
	}
	\subfloat[][\textbf{With CXL}]{
	\label{fig:cxl-architecture}
		\includegraphics[width=0.45\columnwidth]{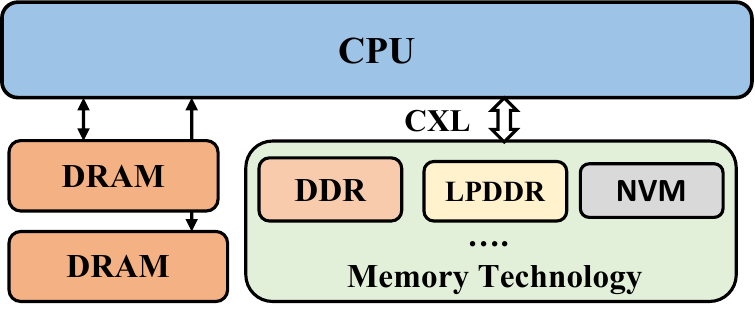}
	}
	\caption{CXL decouples memory from compute.}
	\label{fig:cxl-enabled-memory}
\end{figure}

\begin{figure}[!t]
	\includegraphics[width=0.85\columnwidth]{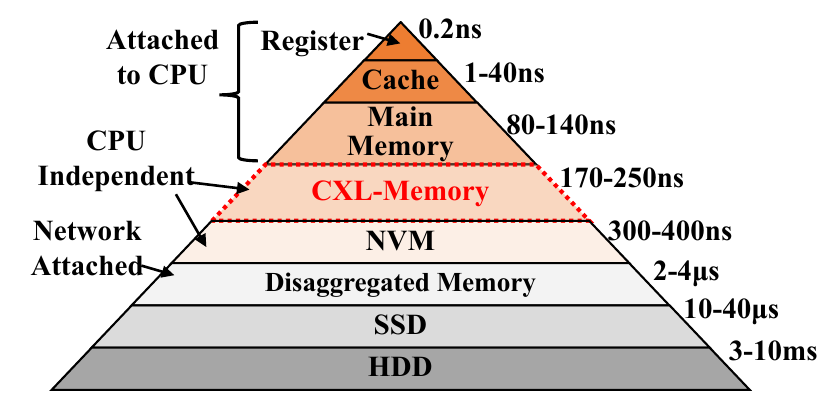}
	\caption{Latency characteristics of memory technologies.}
	\label{fig:memory-hierarchy}
\end{figure}

Compute Express Link (CXL)~\cite{cxl} mitigates this problem by providing an intermediate latency operating point with DRAM-like bandwidth and cache-line granular access semantics.  
CXL protocol allows a new memory bus interface to attach memory to the CPU (Figure~\ref{fig:cxl-enabled-memory}).  
From a software perspective, \cxlmem appears to a system as a CPU-less NUMA node where its memory characteristics (\eg, bandwidth, capacity, generation, technology, \etc) are independent of the memory directly attached to the CPU. 
This allows flexibility in memory subsystem design and fine-grained control over the memory bandwidth and capacity~\cite{next-plarform-cxl, lenovo-cxl, single-socket-cxl}. 
Additionally, as \cxlmem appears like the main memory, it provides opportunities for transparent page placement on the appropriate memory tier.
However, Linux's memory management mechanism is designed for homogeneous CPU-attached DRAM-only systems and performs poorly on \cxlmem system.
In such a system, as memory access latency varies across memory tiers (Figure~\ref{fig:memory-hierarchy}), application performance greatly depends on the fraction of memory served from the fast memory.

To understand whether memory tiering can be beneficial, we need to understand the variety of memory access behavior in existing datacenter applications.
%
For each application, we want to know how much of its memory remains hot, warm, and cold within a certain period and what fraction of its memory is short- vs. long-lived. 
Existing Idle Page Tracking (IPT) based characterization tools~\cite{damon-tool, ipt-tool, idle-mem} do not fit the bill as they require kernel modifications that is often not possible in productions.
Besides, continuous access bit sampling and profiling require excessive CPU and memory overhead.
This may not scale with large working sets. 
Moreover, applications often have different sensitivity towards different types of memory pages (\eg, anon page, file page cache, shared memory, \etc) which existing tools do not account.
To this end, we build \pebstool, a robust and lightweight user-space tool, that uses existing CPU's Precise Event-Based Sampling (PEBS) mechanism to characterize an application's memory access behavior (\S\ref{sec:characterization}).
\pebstool generates a heat-map of memory usage on different types of pages
and provides insights into an application's expected performance with multiple temperature tiers.


We use \pebstool to profile a variety of large memory-bound applications across different service domains running in our production and make the following observations. 
\textbf{(1)} Meaningful portions of application working sets can be warm/cold.
We can offload that to a slow tier memory without significant performance impact. 
\textbf{(2)} A large fraction of anon memory (created for a program's stack, heap, and/or calls to mmap) tends to be hotter, while a large fraction of file-backed memory tends to be relatively colder.
\textbf{(3)} Page access patterns remain relatively stable for meaningful time durations (minutes to hours).
This is enough to observe application behavior and make page placement decisions in kernel-space.
\textbf{(4)} With new (de)allocations, actual physical page addresses can change their behavior from hot to cold and vice versa fairly quickly.
Static page allocations 
can significantly degrade performance.
Considering the above observations, we design an OS-level transparent page placement mechanism -- \TPP, to efficiently place pages in a tiered-memory systems so that relatively hot pages remain in fast memory tier and cold pages are moved to the slow memory tier (\S\ref{sec:tpp}).
\TPP has three prime components: \textbf{(a)} a lightweight reclamation mechanism to demote colder pages to the slow tier node; \textbf{(b)} decoupling the allocation and reclamation logic for multi-NUMA systems to maintain a headroom of free pages on fast tier node; and \textbf{(c)} a reactive page promotion mechanism that efficiently identifies hot pages trapped in the slow memory tier and promote them to the fast memory tier to improve performance.
We also introduce support for page type-aware allocation across the memory tiers -- preferably allocate sensitive anon pages to fast tier and file caches to slow tier.
With this optional application-aware setting, \TPP can act from a better starting point and converge faster for applications with certain access behaviors.

We choose four production workloads that constitute significant portion of our server fleet and run them on a  system that support CXL 1.1 specification (\S\ref{sec:eval}).
We find that \TPP provides the similar performance behavior of all memory served from the fast memory tier.
For some workloads, this holds true even when local DRAM is only 20\% of the total system memory.
\TPP moves all the effective hot memory to the fast  memory tier and improves default Linux's performance by up to 18\%.
We compare \TPP against NUMA Balancing~\cite{autonuma} and AutoTiering~\cite{autotiering}, two state-of-the-art solutions for tiered memory.
\TPP outperforms both of them by 5--17\%.

We make the following contributions in this paper:

\begin{denseitemize}
	\item We present \pebstool, a lightweight user-space memory characterization tool. We use it to understand workload's memory consumption behavior and assess the scope of tiered-memory in hyperscale datacenters (\S\ref{sec:characterization}). We \href{https://github.com/facebookresearch/chameleon}{open source \pebstool}.
	\item We propose \TPP for efficient memory management on a tiered-memory system (\S\ref{sec:tpp}).
	\rev{We publish the \href{https://lwn.net/Articles/876993/}{source code of \TPP}. A major portion of it has been merged to Linux kernel v5.18. Rest of it is under an upstream discussion.} 
	\item We evaluate \TPP on a CXL-enabled tiered-memory systems with real production workloads (\S\ref{sec:eval}) for years. 
	\TPP makes tiered memory systems as performant as an ideal system with all memory in local tier. 
    For datacenter applications, \TPP improves default Linux's performance by up to 18\%.
    It also outperforms NUMA Balancing and AutoTiering by 5--17\%. 
\end{denseitemize}

To the best of our knowledge, we are the first to characterize and evaluate an end-to-end practical \cxlmem system that can be readily deployed in hyperscale datacenters. 

\section{Motivation}

\begin{figure}[h]
	\includegraphics[width=0.9\columnwidth]{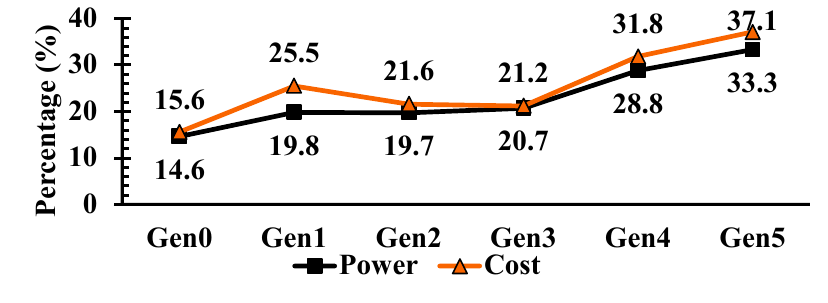}
	\caption{Memory as a percentage of rack TCO and power across different hardware generations of \meta.}
	\label{fig:dram-tco}
\end{figure}

{\bf Increased Memory Demand in Datacenter Applications.} 
To build low-latency services, in-memory computation has become a norm in datacenter applications.
This has led to rapid growth in memory demands across the server fleet. 
With new generations of CPU and DRAM technologies, memory is becoming the more prominent portion of rack-level power and total cost of ownership (TCO). 
(Figure~\ref{fig:dram-tco}).  

\begin{figure}[!t]
		\includegraphics[width=0.9\columnwidth]{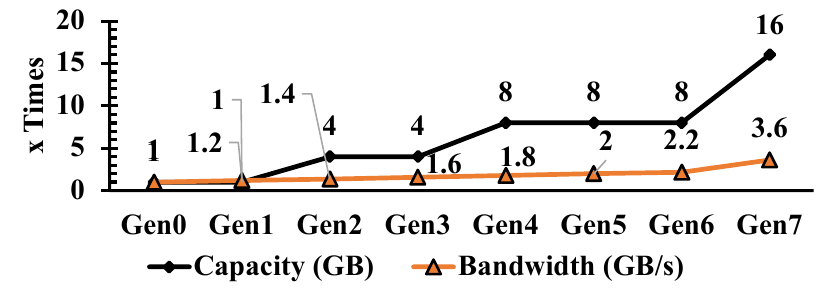}
	\caption{Memory bandwidth and capacity increase over time.}
	\label{fig:dram-demand}
\end{figure}

{\bf Scaling Challenges in Homogeneous Server Designs.} 
In today's server architectures, memory subsystem design is completely dependent on the underlying memory technology support in the CPUs.
This has several limitations: (a) memory controllers only support a single generation of memory technology which limits mix-and-match of different technologies with different cost-per-GB and bandwidth vs. latency profiles; (b) memory capacity comes at power-of-two granularity which limits finer grain memory capacity sizing; (c) there are limited bandwidth vs. capacity points per DRAM generation (Figure~\ref{fig:dram-demand}).
This forces higher memory capacity in order to get more bandwidth on the system. Such tight coupling between CPU and memory subsystem restricts the flexibility in designing efficient memory hierarchies and leads to stranded compute, network, and/or memory resources. Prior bus interfaces that allow memory expansion are also proprietary to some extent~\cite{upi, htx3, infinity} and not commonly supported across all the CPUs~\cite{genz, opencapi, ccix}.
Besides, high latency characteristics and lack of coherency limit their viability in hyperscalers. 
 
\begin{figure}[!t]
	\centering
	\subfloat[][\textbf{Without CXL}]{
	\label{fig:cxl-less-bw}
		\includegraphics[width=0.48\columnwidth]{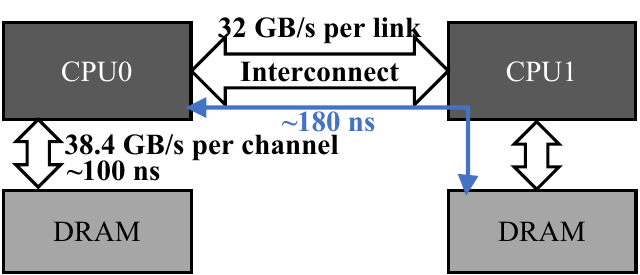}
	}
	\subfloat[][\textbf{With CXL}]{
	\label{fig:cxl-bw}
		\includegraphics[width=0.48\columnwidth]{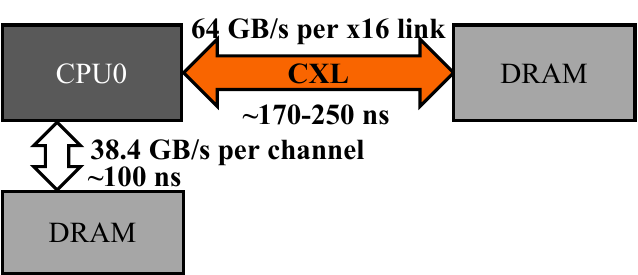}
	}
	\caption{\cxlsystem compared to a dual-socket server.}
	\label{fig:cxl-numa-alike}
\end{figure}

{\bf CXL for Designing Tiered-Memory Systems.}
CXL~\cite{cxl} is an open, industry-supported interconnect based on the PCI Express (PCIe) interface. 
It enables high-speed, low latency communication between the host processor and devices (\eg, accelerators, memory buffers, smart I/O devices, \etc) while expanding memory capacity and bandwidth.  
CXL provides byte addressable memory in the same physical address space and allows transparent memory allocation using standard memory allocation APIs. 
It allows cache-line granularity access to the connected devices and underlying hardware maintains coherency and consistency. 
With PCIe 5.0, CPU to CXL interconnect bandwidth will be 
similar to the cross-socket interconnects (Figure~\ref{fig:cxl-numa-alike}) on a dual-socket machine.
\cxlmem access latency is also similar to the NUMA access latency. CXL adds around 50-100 nanoseconds of extra latency over normal DRAM access.
This NUMA-like behavior with main memory-like access semantics makes \cxlmem a good candidate for the slow-tier in datacenter memory hierarchies.

CXL solutions are being developed and incorporated by leading chip providers~\cite{intel-cxl, amd-cxl, micron-cxl, samsung-cxl, lenovo-cxl, single-socket-cxl}.
All the tools, drivers, and OS changes required to support CXL are open sourced so that anyone can contribute and benefit directly without relying on single supplier solutions.
CXL relaxes most of the memory subsystem limitations mentioned earlier. It enables flexible memory subsystem designs with desired memory bandwidth, capacity, and cost-per-GB ratio based on workload demands.
This helps scale compute and memory resources independently and ensure a better utilization of stranded resources.


{\bf Scope of CXL-Based Tiered-Memory Systems.} 
Datacenter workloads rarely use all of the memory all the time~\cite{googledisagg, google-cluster-trace-2019, alibaba-cluster-trace, snowset, tmo}. 
Often an application allocates a large amount of memory but accesses it infrequently~\cite{googledisagg, memtrade}. 
We characterize four popular applications in our production server fleet and find that 55-80\% of an application's allocated memory remains idle within any two minutes interval (\S\ref{subsec:temperature}). 
Moving this cold memory to a slower memory tier can create space for more hot memory pages to operate on the fast memory tier and improve application-level performance. 
Besides, it also allows reducing TCO by flexible server design with smaller fast memory tier and larger but cheaper slow memory tier.

As CXL-attached slow memory can be of any technology (\eg, DRAM, NVM, LPDRAM, \etc), for the sake of generality, we will call a memory directly attached to a CPU as local memory and a CXL-attached memory as \cxlmem.

{\bf Lightweight Characterization of Datacenter Applications.} 
In a hyperscaler environment, different types of rapidly evolving applications consume a 
production 
server's resources.
\rev{Applications can have different requirements, \eg, some can be extremely latency sensitive, while others can be memory bandwidth sensitive.
Sensitivity towards different memory page types can also vary for different applications.
To understand the scope of tiered-memory system for existing datacenter applications, we need to characterize their memory usage pattern and quantify the opportunity of memory offloading at different memory tiers for different page types. 
This insight can help system admins decide on the flexible and optimized memory configurations to support different workloads.}

Existing memory characterization tools fall short of providing these insights. 
\rev{For example, access bit-based mechanism \cite{damon-tool, ipt-tool, idle-mem} cannot track detailed memory access behavior because it tells whether a given page is accessed within a certain period of time.
Even if a page gets multiple accesses within a tracking cycle, IPT-based tools will count that as a single access.}
Moreover, IPT only provides information in the physical address space -- we cannot track memory allocation/deallocation if a physical page is re-used by multiple virtual pages.
\rev{The overhead of such tools significantly increases with the application's memory footprint size. 
Even for application's with 10s of GB working set size, the overhead of IPT-based tool can be 20--90\%~\cite{daptrace}.}
Similarly, complex PEBS-based user-space tools ~\cite{hemem, atmem, unimem, pebs-hybrid-memory} lead to high CPU overheads (more than 15\% per core) and often slow down the application. None of these tools generate page type-aware heat map.

\section{Characterizing 
Datacenter Applications}
\label{sec:characterization}
\begin{figure*}[t]
	\centering
	\subfloat[][\textbf{\collector}]{
	\label{fig:pebs-collector}
		\includegraphics[width=0.3\textwidth]{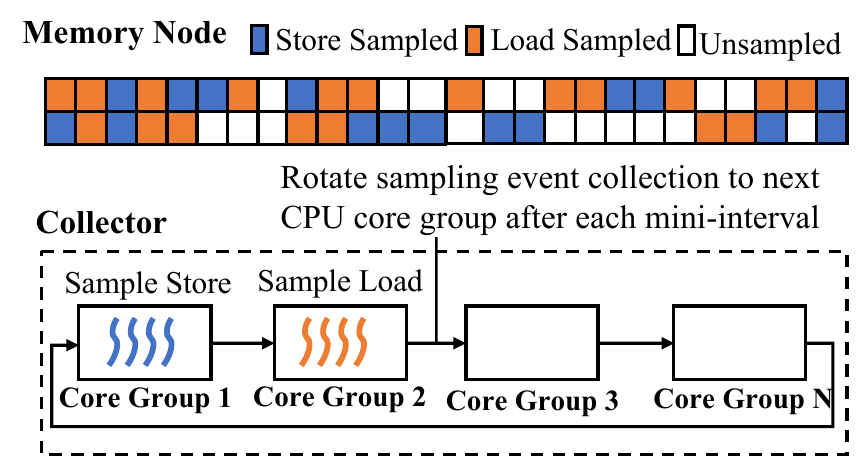}
	}
	\subfloat[][\textbf{\processor}]{
	\label{fig:pebs-processor}
		\includegraphics[width=0.3\textwidth]{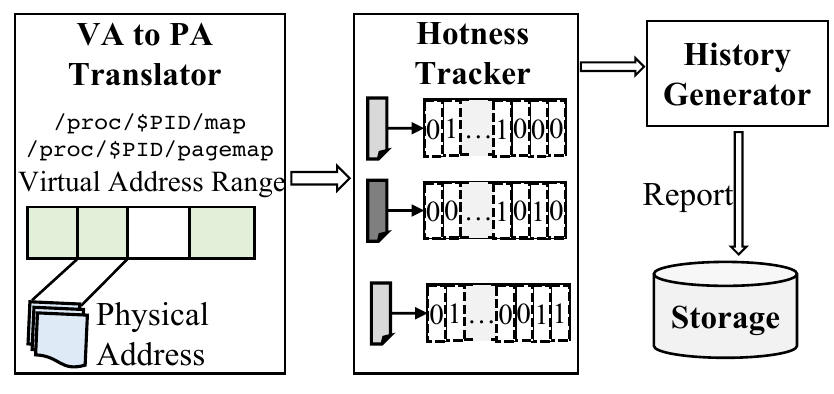}
	}
	\subfloat[][\textbf{Workflow within a Cycle}]{
	\label{fig:pebs-interval}
		\includegraphics[width=0.3\textwidth]{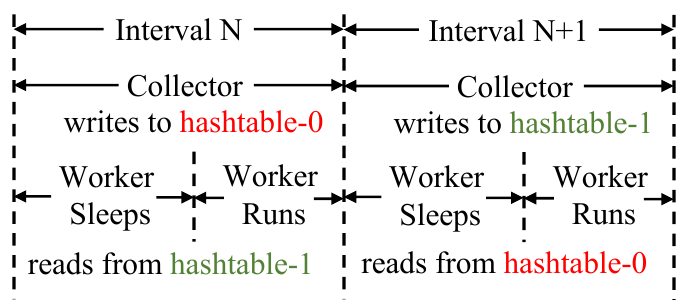}
	}
	\caption{Overview of \pebstool components (left) and workflow (right)}
\end{figure*}


To understand the scope of tiered-memory in hyperscale applications, 
we develop {\pebstool}, a light-weight user-space memory access behavior characterization tool.
\rev{The objective of developing \pebstool is to allow users hop on any existing datacenter production machine and readily deploy it without disrupting the running application(s) and modifying the underline kernel.
\pebstool's overhead needs to be comparatively lower such that it does not notably affect a production application's behavior. 
\pebstool's prime use case is to understand an application's memory access behavior, {\ie}, what fraction of an application's memory remains hot-warm-cold, how long a page survive on a specific temperature tier, how frequently they get accessed and so on.
In practice, to characterize a certain type of application, we expect to run \pebstool for a few hours on a tiny fraction of servers in the whole fleet.}

\rev{Considering the above objectives, we design \pebstool with two primary components -- a \collector and a \processor -- running simultaneously on two different threads.}
\collector utilizes the PEBS mechanism of modern CPUs
to collect hardware-level performance events related to memory accesses.
\processor uses the sampled information to generate insights. 

{\bf \collector.}
\rev{\collector samples last level cache (LLC) misses for demand loads 
(event \texttt{MEM\_LOAD\_RETIRED.L3\_MISS}) 
and optionally TLB misses for demand stores
(event \texttt{MEM\_INST\_RETIRED.STLB\_MISS\_STORES}).} 
Sampled records provide us with the \textsc{PID} and virtual memory address for the memory access events.
Like any other sampling mechanism, accuracy of PEBS depends on the sampling rate -- frequent samples provide higher accuracy.
High sampling rate, however, \rev{incurs higher performance overhead directly on application threads and demands more CPU resources for \pebstool's \processor thread}.
In our fleets, sampling rate is configured as one sample for every 200 events, which appears as a good trade-off between overhead and accuracy.
\rev{Note the choice of the sampling event on the store side is due to hardware limitations, \ie,  there is no precise event for LLC-missed stores, 
likely because stores are marked complete once TLB translation is done in modern high-end CPUs. 
We have conveyed the concern on this limitation to major x86 vendors in multiple occasions.}

\rev{To improve flexibility}, the \collector divides all CPU cores into a set of groups and enables sampling on one or more group(s) at a time (Figure~\ref{fig:pebs-collector}). 
After each \texttt{mini\_interval} (by default, 5 seconds), the sampling thread rotates to the next core group.
This duty-cycling helps further tune the trade-off between overhead and accuracy. 
\rev{It also allows sampling different events on different core groups. 
For example, for latency-critical applications, one can choose to sample half of the cores at a time. 
On the other hand, for store-heavy applications, one can enable load sampling on half of the cores and store sampling on the other half at the same time.}

The \collector reads the sampling buffer and writes into one of the two hash tables.
After each \texttt{interval} (by default, 1 minute), the \collector wakes up the \processor to process data in current hash table and moves to the other hash table for storing next interval's sampled data.

{\bf \processor.}
The \processor (Figure~\ref{fig:pebs-processor}) runs on a separate thread to read page access information and generate insights on memory access behavior. 
 It considers the address of a sampled record as a virtual page access where the page size is defined by the OS.
 This makes it generic to systems with any page granularities (\eg, 4KB base page, 2MB huge page, \etc).
To generate statistics on both virtual- and physical-spaces, the \processor finds the corresponding physical page mapped to the sampled virtual page. 
This address translation can cause high overhead if the target process's working set size is extremely large (\eg, terabyte-scale).
One can configure the \processor to disable the physical address translation and characterize an application only on its virtual-space access pattern.

For each page, a 64-bit bitmap tracks its activeness within an interval.
If a page is active within an interval, the corresponding bit is set. 
At the end of each interval, the bitmap is left-shifted one bit to track for a new interval.
One can use multiple bits for one interval to capture the page access frequency, at the cost of supporting shorter history.
After generating the statistics and reporting them, the \processor sleeps (Figure~\ref{fig:pebs-interval}).
%

To characterize an application deployed on hundreds of thousands of servers, we run \pebstool on a small set of servers only for a few hours.
\rev{During our analysis, we chose servers where system-wide CPU and memory usage does not go above 80\%.
In such production environments, }
we do not notice any service-level performance impact while running \pebstool.
CPU overhead is within 3--5\% of a single core.
\rev{However, on a synthetic workload that is memory bandwidth sensitive and uses all the CPU cores so that \pebstool needs to contend for CPU, we lose 7\% performance due to profiling.}


\subsection{Production Workload Overview}
\label{subsec:characterize}

We use \pebstool to characterize most popular memory-bound applications running for years across our production server fleet serving live traffic on four diverse service domains. 
These workloads constitute a significant portion of the server fleet and represent a wide variety of our workloads~\cite{softsku, accelerometer}.
{\bf Web} implements a Virtual Machine 
to serve web requests. 
Web1 is a HipHop Virtual Machine (HHVM)-based and Web2 is a Python-based service. 
{\bf Cache} is a large distributed-memory object caching service  lying between the web and database tiers for low-latency data-retrieval. 
{\bf Data Warehouse} is a unified computing engine for parallel data processing on compute clusters. 
This service manages and coordinates the execution of long and complex batch queries on data across a cluster.
{\bf Ads} are compute heavy workloads that retrieve in-memory data and perform machine learning computations.

\begin{figure}[!t]
	\includegraphics[width=0.85\columnwidth]{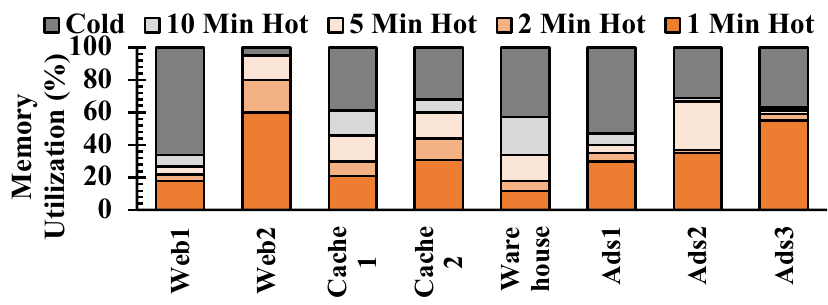}
	\caption{Application memory usage over last N mins.}
	\label{fig:pebs-total-hottness-single}
\end{figure}

\subsection{Page Temperature}
\label{subsec:temperature}
In datacenter applications, a significant amount of allocated memory remains cold beyond a few minutes of intervals (Figure~\ref{fig:pebs-total-hottness-single}).
Web, Cache, and Ads use 95--98\% of the system's total memory capacity, but within a two-minute interval, they use 22--80\% of the total allocated memory on average.


Data Warehouse is a compute-heavy workload where a specific computation operation can span even terabytes of memory. 
This workload consumes almost all the available memory within a server. 
Even here, on average, only 20\% of the accessed memory is hot within a two-minute interval.

    \textbf{Observation:} A significant portion of a datacenter application's accessed memory remain cold for minutes. Tiered memory system can be a good fit for such cold memory if page placement mechanism can move these cold pages to a lower memory tier.

\subsection{Temperature Across Different Page Types}
\label{subsec:temperature-type}
\begin{figure}[!t]
	\includegraphics[width=0.85\columnwidth]{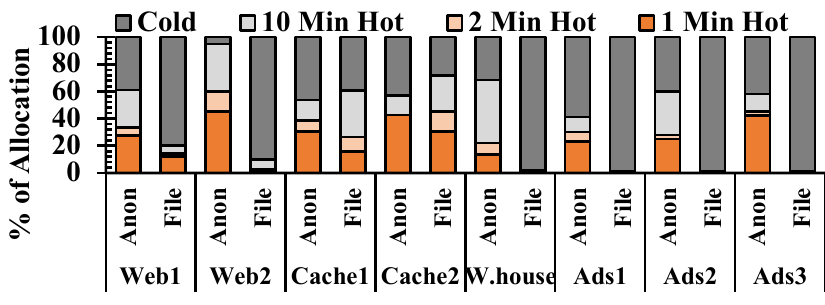}
	\caption{Anon pages tends to be hotter than file pages.}
	\label{fig:pebs-types-hottness-single}
\end{figure}

Applications consume different types of pages based on application logic and execution demand.
However, the fraction of anons (anonymous pages) remain hot is higher than the fraction of files (file pages).
For Web, within a two-minute interval, 35--60\% of the total allocated anons remain hot; for files, in contrast, it is only 3--14\% of the total allocation (Figure~\ref{fig:pebs-types-hottness-single}).

Cache applications use \texttt{tmpfs}~\cite{tmpfs} for a fast in-memory lookup.
Anons are used mostly for processing queries.
As a result, file pages contribute significantly to the total hot memory.
However, for Cache1, 40\% of the anons get accessed within every two minutes, while the fraction for file is only 25\%.
For Cache2, the fraction of anon and file usage is almost equal within a two-minute time window.
However, within a minute interval, even for Cache2, higher fraction of anons (43\%) remain hot over file pages (30\%).

Data Warehouse and Ads use anon pages for computation. 
The file pages are used for writing intermediate computation data to the storage device.
As expected, almost all of hot memories are anons where almost all of the files remain cold.

    \textbf{Observation:} A large fraction of anon pages is hot, while file pages are comparatively colder within short intervals. 

Due to space constraints, we focus on a subset of these applications.
In our analysis, we find the similar behavior from the rest of the applications.

\begin{figure*}[t]
	\centering
	\subfloat[][\textbf{Web1}]{
	\label{fig:pebs-web-memory-single}
		\includegraphics[width=0.23\textwidth]{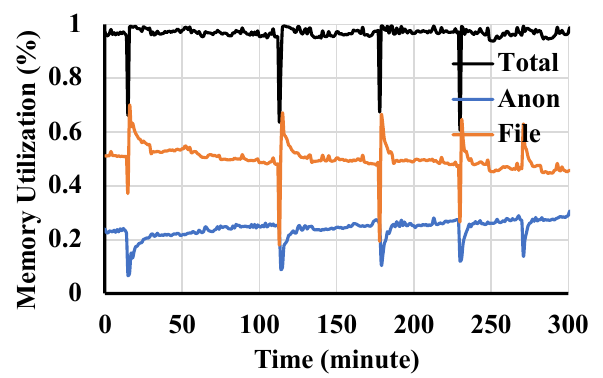}
	}
	\subfloat[][\textbf{Cache1}]{
	\label{fig:pebs-cache1-memory-single}
		\includegraphics[width=0.23\textwidth]{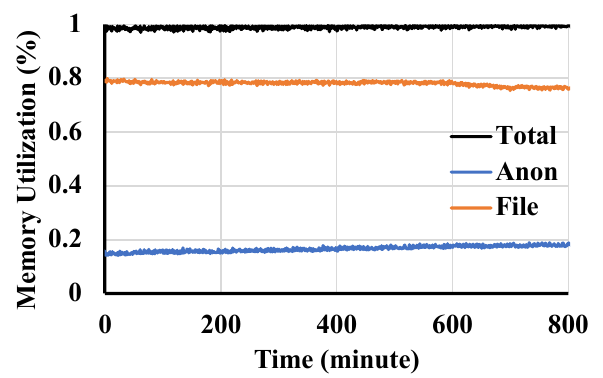}
	}
	\subfloat[][\textbf{Cache2}]{
	\label{fig:pebs-cache2-memory-single}
		\includegraphics[width=0.23\textwidth]{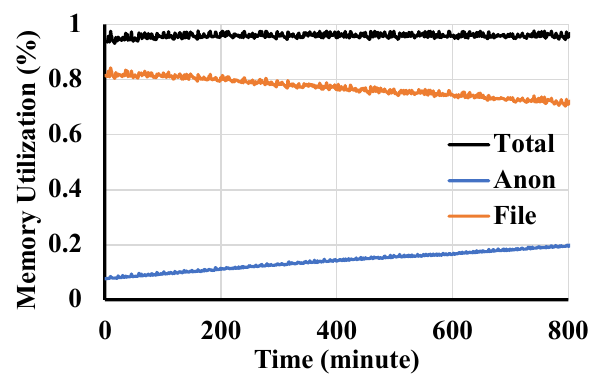}
	}
	\subfloat[][\textbf{Data Warehouse}]{
	\label{fig:pebs-warehouse-memory-single}
		\includegraphics[width=0.23\textwidth]{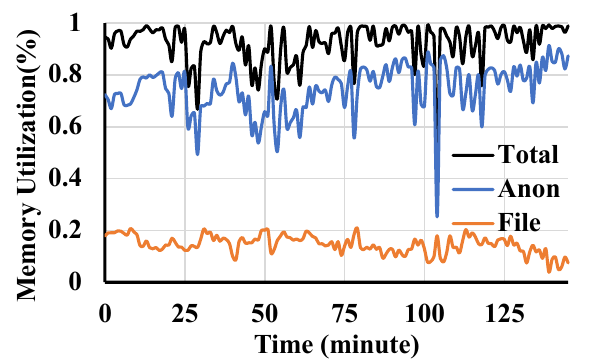}
	}
	\caption{Memory usage over time for different applications}
\end{figure*}

\begin{figure*}[t]
	\centering
	\subfloat[][\textbf{Web1}]{
	\label{fig:pebs-web-qps}
		\includegraphics[width=0.23\textwidth]{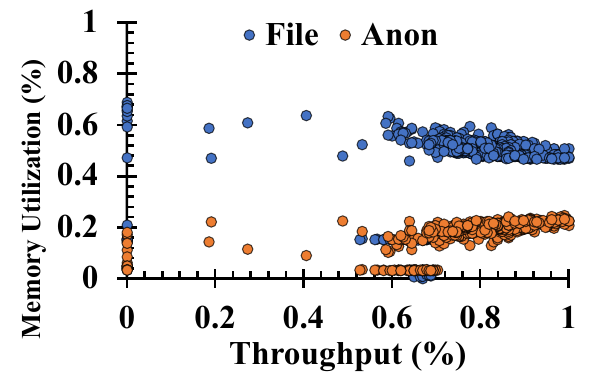}
	}
	\subfloat[][\textbf{Cache1}]{
	\label{fig:pebs-cache1-qps}
		\includegraphics[width=0.23\textwidth]{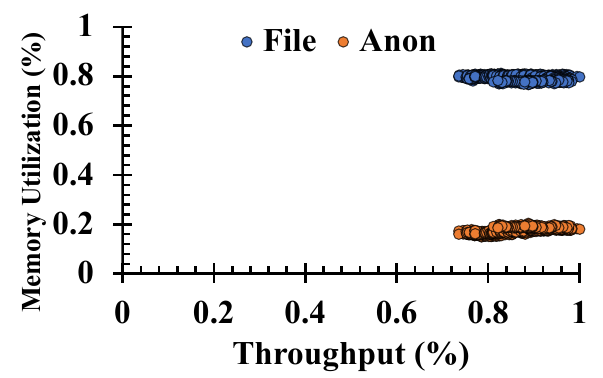}
	}
	\subfloat[][\textbf{Cache2}]{
	\label{fig:pebs-cache2-qps}
		\includegraphics[width=0.23\textwidth]{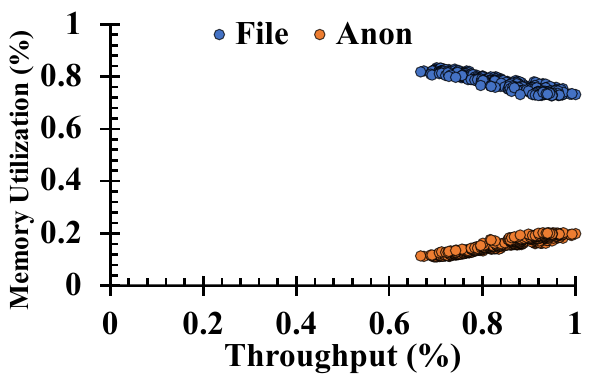}
	}
	\subfloat[][\textbf{Data Warehouse}]{
	\label{fig:pebs-warehouse-qps}
		\includegraphics[width=0.23\textwidth]{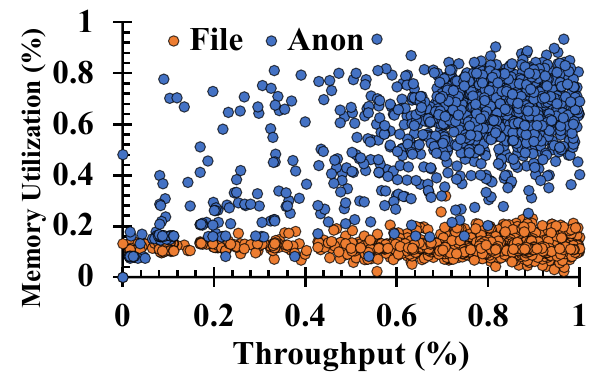}
	}
	\caption{Workloads' sensitivity towards anons and files varies. High memory capacity utilization provides high throughput.}
	\label{fig:pebs-qps-memory}
\end{figure*}

\subsection{Usage of Different Page Types Over Time}
\label{subsec:usage-type}
When the Web service starts, it loads the virtual machine's binary and bytecode files into memory. 
As a result, at the beginning, file caches occupy a significant portion of the memory. 
Overtime, anon usage slowly grows and file caches get discarded to make space for the anon pages (Figure~\ref{fig:pebs-web-memory-single}).

Cache applications mostly use file caches for in-memory look-ups. 
As a result, file pages consume most of the allocated memory.
For Cache1 and Cache2 (Figure~\ref{fig:pebs-cache1-memory-single}-\ref{fig:pebs-cache2-memory-single}), 
the fraction of files hovers around 70--82\%.
While the fraction of anon and file is almost steady, if at any point, anon usage grows, file pages are discarded to accommodate newly allocated anons.

For Data Warehouse workload, anon pages consume most of the allocated memory -- 85\% of the total allocated memory are anons and rest of the 15\% are file pages (Figure~\ref{fig:pebs-warehouse-memory-single}).
The usage of anon and file pages mostly remains steady.

    \textbf{Observation:} Although anon and file usage may vary over time, applications mostly maintain a steady usage pattern. Smart page placement mechanisms should be aware of page type when making placement decisions. 

\subsection{Impact of Page Types on Performance}
\label{subsec:memory-type-impact}
Figure~\ref{fig:pebs-qps-memory} shows what fractions of different page types are used to achieve a certain application-level throughput.
Memory-bound application's throughput improves with high memory utilization.
However, workloads have different levels of sensitivity toward different page types. 
For example, Web's throughput improves with the higher utilization of anon pages (Figure~\ref{fig:pebs-web-qps}).

For Cache, \texttt{tmpfs} is allocated during initialization period.
Besides, Cache1 uses a fixed amount of anons throughout its life cycle.
As a result, we cannot observe any noticeable relation between anon or file usage and the application throughput (Figure~\ref{fig:pebs-cache1-qps}).
However, for Cache2, we can see high throughput is achieved with comparatively higher utilization of anons (Figure~\ref{fig:pebs-cache2-qps}). 
Similarly, Data Warehouse application maintains a fixed amount of file pages. However, it consumes different amount of anons at different execution period and the highest throughput is achieved when the total usage of anons reaches to its highest point (Figure~\ref{fig:pebs-warehouse-qps}).

    \textbf{Observation:} Workloads have different levels of sensitivity toward different page types that varies over time.

\subsection{Page Re-access Time Granularity}
\label{subsec:charcterize-delta}
\begin{figure}[!t]
	\includegraphics[width=0.8\columnwidth]{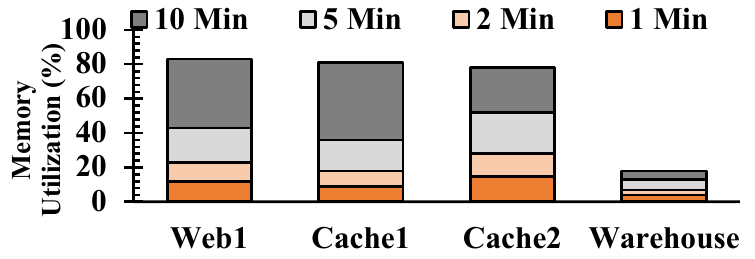}
	\caption{Fraction of pages re-accessed at different intervals.}
	\label{fig:delta-single-socket}
\end{figure}

Cold pages often get re-accessed at a later point.
Figure~\ref{fig:delta-single-socket} shows the fraction of pages that become hot after remaining cold for a certain interval.
For Web, almost 80\% of the pages are re-accessed within a ten-minute interval. 
This indicates Web mostly repurposes pages allocated at an earlier time.
Same goes for Cache -- randomly offloading cold memory can impact performance as a good chunk of colder pages get re-accessed within a ten-minute window.

However, Data Warehouse shows different characteristics. 
For this workload, anons are mostly newly allocated -- within a ten-minute interval, only 20\% of the hot file pages are previously accessed.
Rest of them are newly allocated.

    \textbf{Observation:} Cold page re-access time  varies for workloads. Page placement on a tiered memory system should be aware of this and actively move hot pages to lower memory nodes to avoid high memory access latency.

From above observations, tiered memory subsystems can be a good fit for datacenter applications as there exists significant amount of cold memory with steady access patterns.
\section{Design Principles of \TPP}
\label{sec:tpp-principal}

\rev{
With the advent of CXL technologies, hyperscalers are embracing CXL-enabled heterogeneous tiered-memory system where different memory tier has different performance characteristics~\cite{pond-azure, directcxl}.
For performance optimization in such systems, transparent page placement mechanism (\TPP) is needed to handle pages with varied hotness characteristics on appropriate temperature tiers.
To design \TPP for next-generation tiered-memory systems, we consider following questions:}


\begin{denseitemize}
	\item \rev{What is an ideal layer to implement \TPP functionalities?}
	\item \rev{How to detect page temperature?}
	\item \rev{What abstraction to provide for accessing \cxlmem?}
\end{denseitemize}

\rev{In this section, we discuss the rationale and trade-offs behind design choices for \TPP.}

\textbf{Implementation Layer.}
\rev{Application-transparent page placement mechanism can be employed both in the user- and kernel-space.
In user-space, a \pebstool-like tool can be used to detect page temperatures and perform NUMA migrations using user-space APIs (\eg, \texttt{move\_pages()}). To identify what to migrate, the migration tool needs to implement user-space page lists and history management. This technique entails overheads due to user-space to kernel-space context switching.
It also adds processing overheads due to history management in user space. 
Besides, there are memory overheads due to page information management in user-space that may not scale with large working sets. 
While this is acceptable for profiling tools that run for short intervals on a small sample of servers in the fleet, it can be prohibitively expensive when they run continuously on all production fleet. 
Considering these, we design \TPP as a kernel feature as we think it's less complex to implement and more performant over user-space mechanisms. }

\textbf{Page Temperature Detection.}
\rev{There are several different techniques that can potentially be used for page temperature detection including PEBS, Page Poisoning, and NUMA Balancing. 
PEBS can be utilized in kernel-space to detect page temperature. However, as PEBS counters are not standardized across CPU vendors, a generic precise event-based kernel implementation for page temperature detection that works across all hardware platforms is not feasible.
Additionally, limited number of perf counters are supported in CPUs and are generally required to be exposed in user-space.
More importantly, as mentioned earlier, even with optimizations, PEBS-based profiling is not good enough as an always-running component of \TPP for high pressure workloads.}

\rev{Sampling and poisoning a few pages within a memory region to track their access events is another well-established approach~\cite{thermostat, damon-reclaim, autonuma, ipt} for finding hot/cold pages.
To detect a page access, IPT-based~\cite{ipt, damon-reclaim} approach needs to clear the page's  \texttt{accessed bit} and flush the corresponding TLB entry.
This requires monitoring \texttt{accessed bits} at high frequency which results in unacceptable slowdowns~\cite{thermostat, damon-reclaim}.
Thermostat~\cite{thermostat} solves this problem by sampling at 2MB page granularity which makes it effective specifically for huge-pages.
One of our design goals behind \TPP is that it should be agnostic to page size.
In our production environment, 
application owners generally pre-allocate 2MB and 1GB pages 
and use them to allocate text regions (code), static data structures, and slab pools that serve request allocations. 
In most cases, these large pages are hot and should never get demoted to \cxlmem.} 

\rev{NUMA Balancing (also known as AutoNUMA)~\cite{autonuma} is transparent to OS page sizes.
It generates a minor page fault when the sampled page gets accessed.
Periodically incurring page faults on most frequently accessed pages can lead to high overheads. To address this, when designing \TPP, we chose to only leverage minor page fault as a temperature detection mechanism for \cxlmem. 
As \cxlmem is expected to hold warm and cold pages, this will keep the overhead of temperature detection low. 
For cold page detection in local memory node, we find Linux's existing LRU-based age management mechanism is lightweight and quite efficient.}

\rev{Empirically, we do not see any potential benefit to leverage a more sophisticated page temperature detection mechanism. 
In our experiments (\S\ref{sec:eval}), using kernel LRUs for on-the-fly profiling works well.
Combining LRUs and NUMA Balancing, we can detect most hot pages in \cxlmem at virtually zero overhead as presented in 
Figure~\ref{fig:tpp-wo-constraint} and Table~\ref{tab:tpp-effectiveness}.}

\textbf{Memory Abstraction for \cxlmem.} \rev{One can use \cxlmem as a swap-space to host colder memory using existing in-memory swapping mechanisms~\cite{zswap,frontswap,zram,gcpswap}.
TMO~\cite{tmo} is one such swap-based mechanism that detects cold pages in local memory and moves them to swap-space that is referred to as (z)swap pool. 
However, we do not plan to use \cxlmem as an in-memory swap device.
In such a case, we will effectively lose CXL's most important feature, \ie, load/store access semantics at cache-line granularity.
With swap abstraction, every access to a (z)swapped page will incur a major page fault and we have to read the whole page from \cxlmem.
This will significantly increase the effective access latency far over 200ns and make \cxlmem less attractive.
We chose to build \TPP so that applications can leverage load-store semantics when accessing warm or cold data from \cxlmem. 
}

\rev{While the swap semantics of TMO are not desirable in \cxlmem page placement context, the memory saving through feedback-driven reclamation is still valuable. We think TMO as an orthogonal and complimentary tool to \TPP as it operates one layer above \TPP (\S\ref{subsubsec:tpp-tmo}). TMO runs in user-space and keeps pushing for memory reclamation, while \TPP runs in kernel-space and optimizes page placement for allocated memory between local and \cxlmem tiers. 
}




\section{\TPP for \cxlmem}
\label{sec:tpp}
An effective page placement mechanism should efficiently offload cold pages to slower \cxlmem while aptly identify trapped hot pages in \remote and promote them to the fast memory tier.  
As \cxlmem is CPU-less and independent of the CPU-attached memory, it should be flexible enough to support heterogeneous memory technologies with varied characteristics.
Page allocation to a NUMA node should not frequently halt due to the slower reclamation mechanism to free up spaces.
Besides, an effective policy should be aware of an application's sensitivity toward different page types.

Considering the datacenter workload characteristics and our design objectives, we propose \TPP -- a smart OS-managed mechanism for tiered-memory system.
\TPP places  `hotter' pages in local memory and moves `colder' pages in \cxlmem. 
%
%
\TPP's design-space can be divided across four main areas -- {\bf (a)} lightweight demotion to \cxlmem, {\bf (b)} decoupled allocation and reclamation paths, {\bf (c)} hot-page promotion to \local nodes, and {\bf (d)} page type-aware memory allocation. 


\subsection{Migration for Lightweight Reclamation} 
Linux tries to allocate a page to the memory node local to a CPU where the process is running.
When a CPU's \local memory node fills up, default reclamation pages-out to swap device.  
In such a case, in a NUMA system, new allocations to \local node halts and takes place on \remote until enough pages are freed up.
The slower the reclamation is, the more pages end up being allocated to the \remote.
Besides, invoking paging events in the critical path worsens the average page access latency and impacts application performance.

\begin{figure}[h]
	\includegraphics[width=0.9\columnwidth]{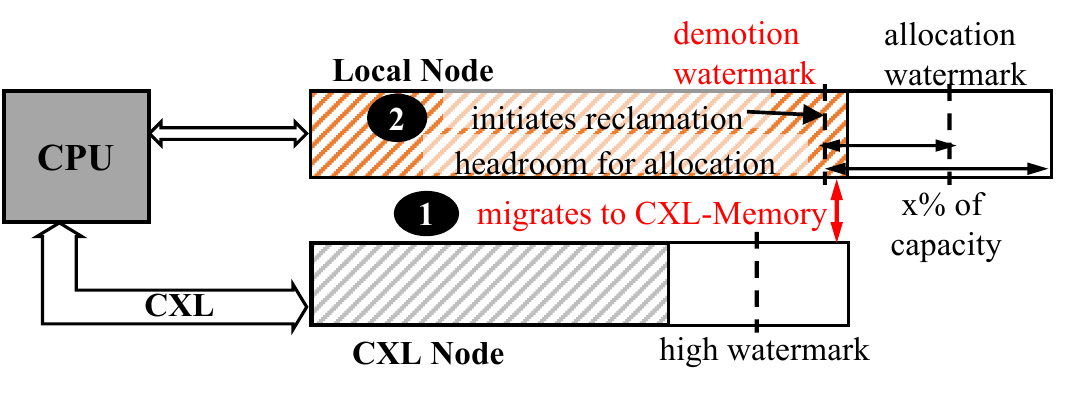}
	\caption{\TPP decouples the allocation and reclamation logics for local memory node. It uses migration for demotion.}
	\label{fig:tpp-demotion}
\end{figure}

To enable a light-weight page reclamation for \local nodes, after finding the reclamation-candidates, instead of invoking swapping mechanism, we put them in to a separate demotion list and try to migrate them to the \remote asynchronously (\circled{1} in Figure~\ref{fig:tpp-demotion}).
Migration to a NUMA node is orders of magnitude faster than swapping.
We use Linux's default LRU-based mechanism to select demotion candidates.
However, unlike swapping, as demoted pages are still available in-memory, along with inactive file pages, we scan inactive anon pages for reclamation candidate selection.
As we start with the inactive pages, chances of hot pages being migrated to \remote during reclamation is very low unless the \local node's capacity is smaller than the hot portion of working set size.
If a migration during demotion fails (\eg, due to low memory on the \remote), we fall back to the default reclamation mechanism for that failed page.
As allocation on \remote is not performance critical, {\remote}s use the default reclamation mechanism (\eg, pages out to the swap device).

If there are multiple {\remote}s, the demotion target is chosen based on the node distances from the CPU.
Although other complex algorithms can be employed to dynamically choose the demotion target based on a {\remote}'s state, this simple distance-based static mechanism turns out to be effective.

\subsection{Decoupling Allocation and Reclamation}
\label{subsec:tpp-decoupling}
Linux maintains three watermarks (\texttt{min}, \texttt{low}, \texttt{high}) for each memory zone within a node.
If the total number of free pages for a node goes below \texttt{low\_watermark}, Linux considers the node is under memory pressure and initiates page reclamation for that node.
In our case, \TPP demotes them to \remote.
New allocation to local node halts till the reclaimer frees up enough memory to satisfy the \texttt{high\_watermark}.
With high allocation rate, reclamation may fail to keep up as it is slower than allocation.
Memory retrieved by the reclaimer may fill up soon to satisfy allocation requests.
As a result, local memory allocations halt frequently, more pages end up in \remote which eventually degrades application performance.

In a multi-NUMA system with severe memory constraint, we should proactively maintain a reasonable amount of free memory headroom on the \local node. 
This helps in two ways. 
First, new allocation bursts (that are often related to request processing and, therefore, both short-lived  and hot) can be directly mapped to the \local node.
Second, \local node can accept promotions of trapped hot pages on {\remote}s. 

To achieve that, we decouple the logic of `reclamation stop' and `new allocation happen' mechanism.
We continue the asynchronous background reclamation process on \local node until its total number of free pages reaches \texttt{demotion\_watermark}, while new allocation can happen if the free page count satisfies a different watermark -- \texttt{allocation\_watermark} (\circled{2} in Figure~\ref{fig:tpp-demotion}). 
Note that demotion watermark is always set to a higher value above the allocation and low watermark so that we always reclaim more to maintain the free memory headroom.

\rev{How 
aggressively one needs to reclaim often depends on the application behavior and available resources. 
For example, if an application has high page allocation demand, but a large fraction of its memory is infrequently accessed, aggressive reclamation can help maintain free memory headroom.
On the other hand, if the amount of frequently accessed pages is larger than the local node's capacity, aggressive reclamation will thrash hot memory across NUMA nodes.} 
Considering these, to tune the aggressiveness of the reclamation process on \local nodes, we provide a user-space \texttt{sysctl} interface (\texttt{/proc/sys/vm/demote\_scale\_factor}) 
to control the free memory threshold for triggering the reclamation on \local nodes. 
By default, its value is \rev{empirically} set to 2\% that means reclamation starts as soon as only a 2\% of the \local node's capacity is available to consume.
\rev{One can use workload monitoring tools~\cite{tmo} to dynamically adjust this value.}

\subsection{Page Promotion from CXL-Node}
\label{tpp:promotion} 
Due to increased memory pressure on \local nodes, new pages may often get allocated to {\remote}s.
Besides, demoted pages may also become hot later as discussed in \S\ref{subsec:charcterize-delta}.
Without any promotion mechanism, hot pages will always be trapped in {\remote}s and hurt application performance.
To promote such pages, we augment Linux's NUMA Balancing~\cite{autonuma}.

{\bf NUMA Balancing for CXL-Memory.}
In NUMA Balancing, a kernel task routinely samples a subset of a process's memory (by default, 256MB of pages) on each memory node.
When a CPU accesses a sampled page, a minor page-fault is generated (known as NUMA hint fault). 
Pages that are accessed from a remote CPU are migrated to that CPU's local memory node (known as promotion).
In a \cxlsystem, it is not reasonable to promote a \local node's hot memory to other \local or {\remote}s. 
Besides, as sampling pages to find a \local node's hot memory generates unnecessary NUMA hint fault overheads,
we limit sampling only to {\remote}s. 

While promoting a {\remote}'s trapped hot pages, we ignore the allocation watermark checking for the target \local node.
This creates more memory pressure to initiate the reclamation of comparatively colder pages on that node. 
If a system has multiple \local nodes, we select the node where the task is running.
When applications share multiple memory nodes, we choose \local node with the lowest memory pressure.


\begin{figure}[!t]
	\includegraphics[width=0.9\columnwidth]{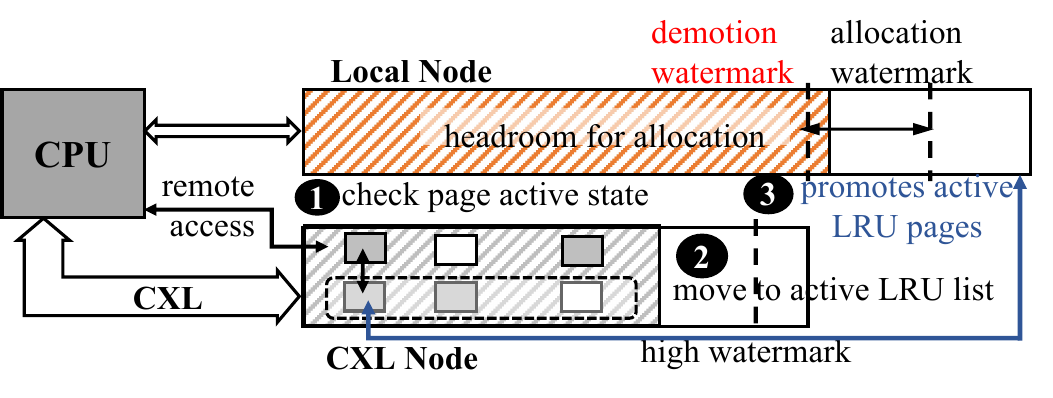}
	\caption{\TPP promotes a page considering its activity state.}
	\label{fig:tpp-promotion}
\end{figure}

{\bf Ping-Pong due to Opportunistic Promotion.}    
When a NUMA hint fault happens on a page, default NUMA balancing instantly promotes the page without checking its active state. 
As a result, pages with very infrequent accesses can still be promoted to the \local node.
Once promoted, these type of pages may shortly become the demotion candidate if the \local nodes are always under pressure. 
Thus, promotion traffic generated from infrequently accessed pages can easily fill up the \local node's free space and generate a higher demotion traffic for {\remote}s. 
This unnecessary traffic can negatively impact an application's performance.
    
{\bf Apt Identification of Trapped Hot Pages.}
To solve this ping-pong issue, instead of instant promotion, we check a page's age through its position in the LRU list maintained by the OS. 
If the faulted page is in inactive LRU, we do not consider the page for promotion  instantly as it might be an infrequently accessed page. 
We consider the faulted page as a promotion candidate only if it is found in the active LRUs (\circled{1} in Figure~\ref{fig:tpp-promotion}).
This significantly reduces the promotion traffic.

However, OS uses LRU lists for reclamation.
If a memory node is not under pressure and reclamation does not kick in, then pages in inactive LRU list do not automatically move to the active LRU list. 
As {\remote}s may not always be under pressure, faulted pages may often be found in the inactive LRU list and, therefore, bypass the promotion filter. 
To address this, whenever we find a faulted page on the inactive LRU list, we mark the page as accessed and move it to the active LRU list immediately (\circled{2} in Figure~\ref{fig:tpp-promotion}). 
If the page still remains hot during the next NUMA hint fault, it will be in the active LRU, and promoted to the \local node (\circled{3} in Figure~\ref{fig:tpp-promotion}).

This helps \TPP add some hysteresis to page promotion. 
Besides, as Linux maintains separate LRU lists for anon and file pages, different page types have different promotion rates based on their respective LRU sizes and activeness. 
This
speeds up the convergence of hot pages across memory nodes.

\begin{figure*}[!t]
	\centering
	\subfloat[][\textbf{Web1}]{
	\label{fig:tpp-web}
		\includegraphics[width=0.24\textwidth]{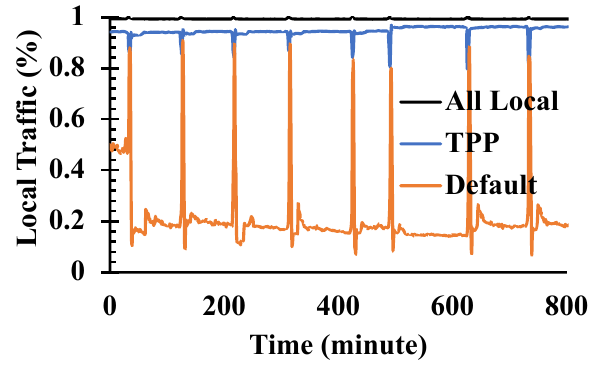}
	}
	\subfloat[][\textbf{Cache1}]{
	\label{fig:tpp-cache1}
		\includegraphics[width=0.24\textwidth]{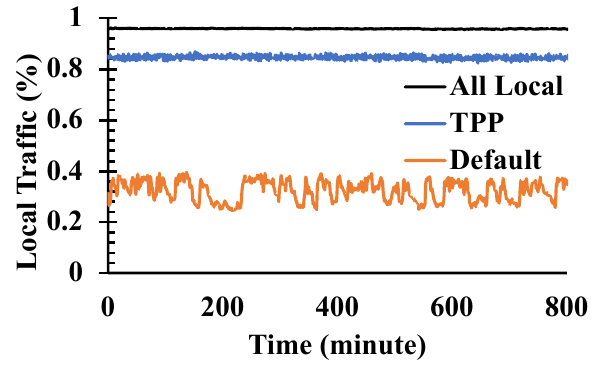}
	}
	\subfloat[][\textbf{Cache2}]{
	\label{fig:tpp-cache2}
		\includegraphics[width=0.24\textwidth]{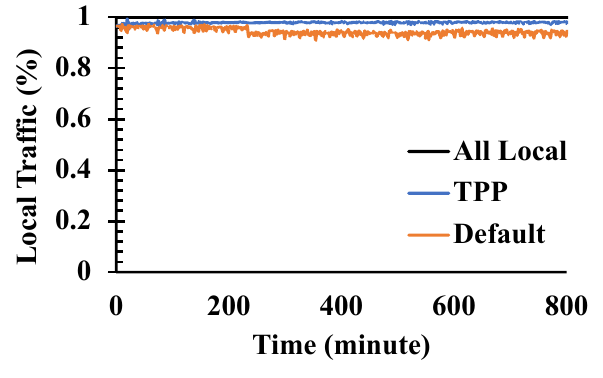}
	}
	\subfloat[][\textbf{Data Warehouse}]{
	\label{fig:tpp-warehouse}
		\includegraphics[width=0.24\textwidth]{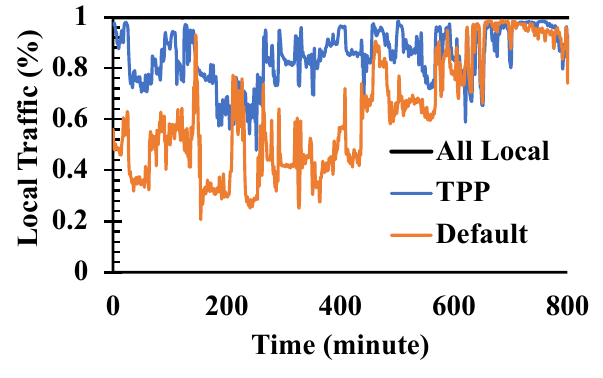}
	}
	\caption{Fast cold page demotion and effective hot page promotion allow \TPP to serve most of the traffics from the \local node.}
	\label{fig:tpp-wo-constraint}
\end{figure*}

\subsection{Page Type-Aware Allocation}
The page placement mechanism we described above is generic to all page types.
However, some applications can further benefit from page type-aware allocation policy.  
For example, production applications often perform lots of file I/O during the warm up phase and generate file caches 
that are accessed infrequently.
As a result, cold file caches eventually end up on {\remote}s.
Not only that, \local memory node being occupied by the inactive file caches often forces anons to be allocated on the {\remote}s that may need to be promoted back later.

To resolve these unnecessary page migrations, we allow an application allocating caches (\eg, file cache, tmpfs, \etc) to the {\remote}s preferrably, while preserving the allocation policy for anon pages.
When this allocation policy is enabled, page caches generated at any point of an application's life cycle will be initially allocated to the \remote.
If a page cache becomes hot enough to be selected as a promotion candidate, it will be eventually promoted to the \local node.
This policy helps applications with infrequent cache accesses run on a system with a small amount of \local memory and large but cheap \cxlmem while maintaining the performance.

\subsection{Observability into \TPP to Assess Performance}

Promotion- and demotion-related statistics can help better understand the effectiveness of the page placement mechanism and debug issues in production environments. 
To this end, we introduce multiple counters to track different demotion and promotion related events and make them available in the user-space through \texttt{/proc/vmstat} interface.

To characterize the demotion mechanism, we introduce counters 
to track the distribution of successfully demoted anon and file pages.
To understand the promotion behavior, we add new counters to collect information on the numbers of sampled pages, the number of promotion attempts, and the number of successfully promoted pages for each of the memory types.

To track the ping-pong issue mentioned in \S\ref{tpp:promotion}, we utilize the unused \texttt{0x40} bit in the page flag to introduce \texttt{PG\_demoted} flag for demoted pages.
Whenever a page is demoted its \texttt{PG\_demoted} bit is set and gets cleared upon promotion.
We also count the number of demoted pages that become promotion candidates.
High value of this counter means \TPP is thrashing across NUMA nodes. 
We add counters for each of the promotion failure scenario (\eg, \local node having low memory, abnormal page references, system-wide low memory, \etc) to reason about where and how promotion fails.

\section{Evaluation}
\label{sec:eval}
We integrate \TPP on Linux kernel v5.12.
We evaluate \TPP with a subset of representable production applications mentioned in \S\ref{subsec:characterize} serving live traffic on tiered memory systems across our server fleet.
We explore the following questions:

\begin{denseitemize}
	\item How effective \TPP is in distributing pages across memory tiers and improving application performance? (\S\ref{subsec:tpp-effectiveness})
	\item What are the impacts of \TPP components? (\S\ref{subsec:tpp-components})
	\item How it performs against state-of-the-art solutions? (\S\ref{subsec:tpp-counterparts})
\end{denseitemize} 

\begin{table}[]
\caption{\TPP is effective over its counterparts. It reduces memory access latency and improves application throughput.}
\resizebox{0.9\columnwidth}{!}{
\begin{tabular}{|c|c|c|c|c|}
\hline
 &  &  &  &  \\
\multirow{-2}{*}{\begin{tabular}[c]{@{}c@{}}Workload/Throughput (\%)\\ (normalized to Baseline)\end{tabular}} & \multirow{-2}{*}{\begin{tabular}[c]{@{}c@{}}Default\\  Linux\end{tabular}} & \multirow{-2}{*}{\TPP} & \multirow{-2}{*}{\begin{tabular}[c]{@{}c@{}}NUMA \\ Balancing\end{tabular}} & \multirow{-2}{*}{AutoTiering} \\ \hline
Web1 (2:1) & 83.5 & \textbf{99.5} & 82.8 & 87.0 \\ \hline
Cache1 (2:1) & 97.0 & \textbf{99.9} & 93.7 & 92.5 \\ \hline
Cache1 (1:4) & 86.0 & \textbf{99.5} & 90.0 & {\color[HTML]{FE0000} Fails} \\ \hline
Cache2 (2:1) & 98.0 & \textbf{99.6} & 94.2 & 94.6 \\ \hline
Cache2 (1:4) & 82.0 & \textbf{95.0} & 78.0 & {\color[HTML]{FE0000} Fails} \\ \hline
Data Warehouse (2:1) & 99.3 & \textbf{99.5} & -- & -- \\ \hline
\end{tabular}
}
\label{tab:tpp-effectiveness}
\end{table}

\rev{For each experiment, we use application-level throughput as the metric for performance.}
\rev{In addition, we use the fraction of memory accesses served from the local node as the key low-level supporting metric.}
We compare \TPP against default Linux (\S\ref{subsec:tpp-effectiveness}), default NUMA Balancing~\cite{autonuma}, AutoTiering~\cite{autotiering}, and TMO~\cite{tmo} (\S\ref{subsec:tpp-counterparts}) (Table~\ref{tab:tpp-effectiveness}).
None of our experiments involve swapping to disks, the whole system has enough memory to support the workload.
We use the default {\it local-node first, then \remote} allocation policy for Linux.

{\bf Experimental Setup.}
We deploy a number of pre-production x86 CPUs with FPGA-based CXL-Memory expansion card that support CXL 1.1 specification.
Memory attached to the expansion card shows up to the OS as a CPU-less NUMA node.
Although our current FPGA-based CXL cards have around 250ns higher latency than our eventual target, we use them for the functional validation. 

{\it We have confirmation from two major x86 CPU vendors that the access latency to \cxlmem is similar to the remote latency on a dual-socket system.}
For performance evaluation, we primarily use dual-socket systems and configure them to mimic our target CXL-enabled system's characteristics (one memory node with all active CPU cores and one CPU-less memory node) according to the guidance of our CPU vendors.
For baseline, we disable the memory node and CPU cores on a socket while enabling sufficient memory on another socket.
Here, single memory node serves the whole working set.

We use two memory configurations where the ratio of \local node and \cxlmem capacity is {\bf (a)} 2:1 (\S\ref{subsubsec:no-constarint}) and {\bf (b)} 1:4 (\S\ref{subsubsec:constarint}).
Configuration {\bf (a)} is similar to our current CXL-enabled production systems where \local node is supposed to serve all the hot working set.
We use configuration {\bf (b)} to stress-test \TPP on a constrained memory scenario -- only a fraction of the total hot working set can fit on the \local node and hot pages are forced to move to the {\remote}.


\subsection{Effectiveness of \TPP}
\label{subsec:tpp-effectiveness}



\subsubsection{Default Production Environment (2:1 Configuration).}
\label{subsubsec:no-constarint}
\textbf{Web.} Web1 performs lots of file I/O during initialization and fills up the \local node.
Default Linux is $44\times$ slower than \TPP during freeing up the \local node.
As a result, new allocation to the \local node halts and anons get allocated to the {\remote} and stay there forever. 
In default Linux, on average, only 22\% of total memory accesses are served from the {\local} node (Figure~\ref{fig:tpp-web}).
As a result, throughput drops by 16.5\%.

Active and faster demotion helps \TPP move colder file pages to {\remote} and allow more anon pages to be allocated in the \local node.
Here, 92\% of the total anon pages are served from the \local node.
As a result, \local node serves 90\% of total memory accesses.
Throughput drop is only 0.5\%.

{\bf Cache.}
Cache applications maintain a steady ratio of anon and file pages throughout their life-cycle.
Almost all the anon pages get allocated to the \local node from the beginning and served from there.
For Cache1, with default Linux, only 8\% of the total hot pages are trapped in \remote. 
As a result, application performance remains very close to the baseline -- performance regression is only 3\%.
\TPP can even minimize this performance gap (throughput is 99.9\% of the baseline) by promoting all the trapped hot files and improving the fraction of traffic served from the \local node (Figure~\ref{fig:tpp-cache1}).

Although most of the Cache2's anon pages reside in the \local node on a default Linux kernel, all of them are not always hot -- only 75\% of the total anon pages remain hot within a two-minute interval.
\TPP can efficiently detect the cold anon pages and demote 
them to the {\remote}.
This allows the promotion of more hot file pages.
On default Linux, \local node serves 78\% of the memory accesses (Figure~\ref{fig:tpp-cache2}) 
and cause 2\% throughput regression. 
\TPP improves the fraction of \local traffic to 91\%. Throughput regression is only 0.4\%.

\begin{figure}[!t]
	\centering
	\subfloat[][\textbf{Cache1}]{
	\label{fig:tpp-cache1-constraint}
		\includegraphics[width=0.48\columnwidth]{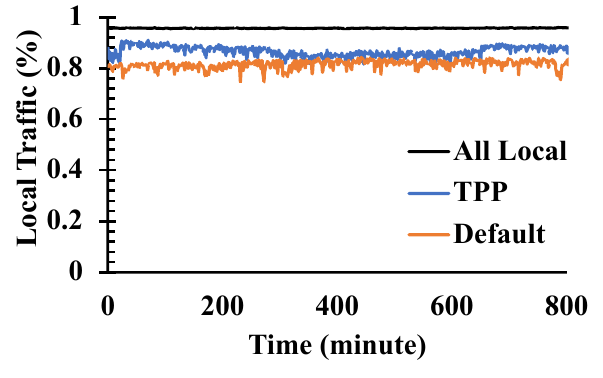}
	}
	\subfloat[][\textbf{Cache2}]{
	\label{fig:tpp-cache2-constraint}
		\includegraphics[width=0.48\columnwidth]{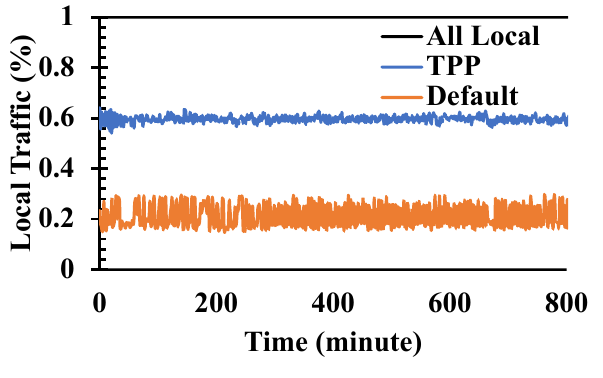}
	}
	\caption{Effectiveness of \TPP under memory constraint.}
	\label{fig:tpp-with-constraint}
\end{figure}

{\bf Data Warehouse.}
This workload generates file caches to store the intermediate processing data.
File caches mostly remain cold.
Besides, only one third of the total anon pages remain hot.
Our default production configuration is good enough to serve all the hot memory from the \local node.
Default Linux and \TPP perform the same (0.5--0.7\% throughput drop).

\TPP improves the fraction of \local traffic by moving relatively hotter anon pages to the \local node.
With \TPP, 94\% of the total anon pages reside on the \local node, while the default kernel hosts only 67\%.
To make space for the hot anon pages, cold file pages are demoted to the {\remote}.
\TPP allows 4\% higher \local node memory accesses over Linux (Figure~\ref{fig:tpp-warehouse}).
 
\subsubsection{Large Memory Expansion with CXL (1:4 Configuration).}
\label{subsubsec:constarint}

Extreme setups like 1:4 configuration allow flexible server design with DRAM as a small-sized \local node and \cxlmem as a large but cheaper memory.
As in our production, such a configuration is impractical for Web and Data Warehouse, we limit our discussion to the Cache applications.
Note that \TPP is effective even for Web and Data Warehouse in such a setup and performs very close to the baseline.

{\bf Cache1.}
In a 1:4 configuration, on a default Linux kernel, file pages consume almost all the \local node's capacity. 
85\% of the total anon pages get trapped to the {\remote} and
throughput drops by 14\%. 
Because of the apt promotion, \TPP can move almost all the {\remote}'s hot anon pages (97\% of the total hot anon pages within a minute) to the \local node.
This forces less latency-sensitive file pages to be demoted to the {\remote}.
%
Eventually, \TPP stabilizes the traffic between the two nodes and \local node serves 85\% of the total memory accesses.
This helps Cache1 achieve the baseline performance -- even though the \local node's capacity is only 20\% of the working set size, throughput regression is only 0.5\%~(Figure~\ref{fig:tpp-cache1-constraint}).

\begin{figure}[!t]
	\centering
	\subfloat[][\textbf{Avg. Memory Access Latency}]{
	\label{fig:tpp-sweep-latency}
		\includegraphics[width=0.48\columnwidth]{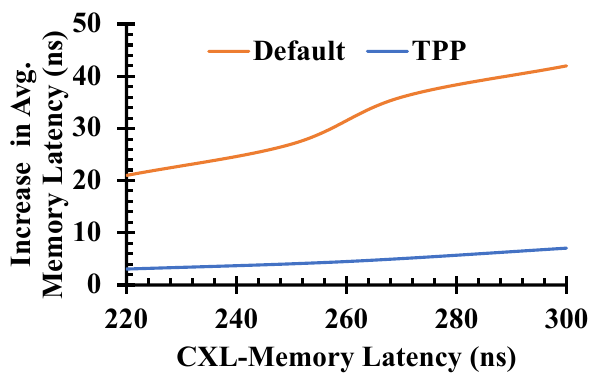}
	}
	\subfloat[][\textbf{Impact on Throughput}]{
	\label{fig:tpp-sweep-throughput}
		\includegraphics[width=0.48\columnwidth]{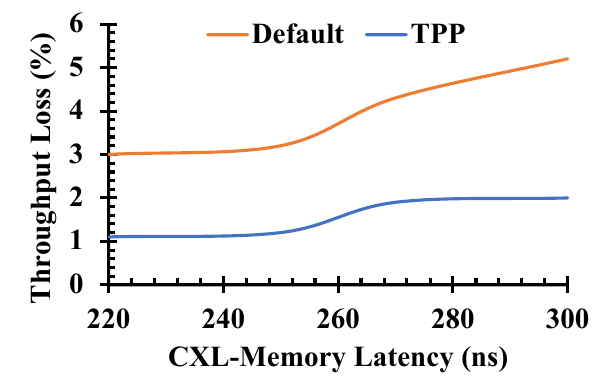}
	}
	\caption{\TPP benefits \remote with varied latency traits.}
	\label{fig:tpp-latency-sweep}
\end{figure}

{\bf Cache2.}
Similar to Cache1, on default Linux, Cache2 experiences 18\% throughput loss.
Only 14\% of the anon pages remain on the \local node~(Figure~\ref{fig:tpp-cache2-constraint}).
 \TPP can bring back almost all the remote hot anon pages (80\% of the total anon pages) to \local node while demoting the cold ones to the \remote.
As Cache2 accesses lots of caches, and caches are now mostly in {\remote}, 41\% of the memory traffic comes from the {\remote}.
Yet, throughput drop is only 5\% with \TPP.

\begin{figure}[!t]
	\centering
	\subfloat[][\textbf{New Allocation}]{
	\label{fig:tpp-cache1-wo-decoupling}
		\includegraphics[width=0.48\columnwidth]{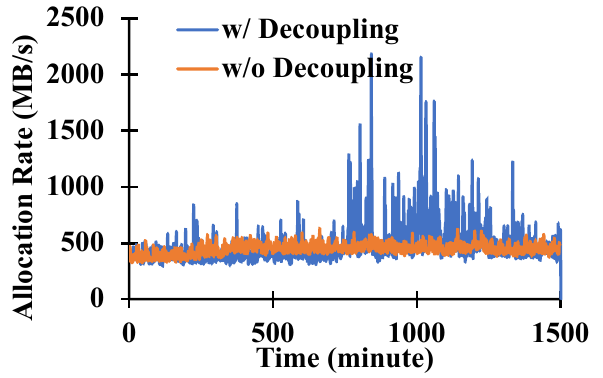}
	}
	\subfloat[][\textbf{Promotion to Toptier Node}]{
	\label{fig:tpp-cache1-w-decoupling}
		\includegraphics[width=0.48\columnwidth]{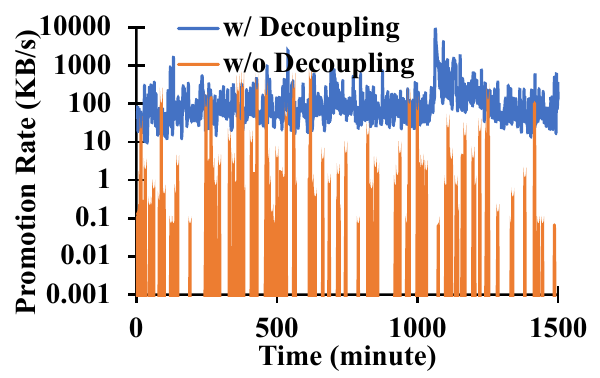}
	}
	\caption{Impact of decoupling allocation and reclamation.}
	\label{fig:tpp-component-decoupling}
\end{figure}

\subsubsection{\TPP with Varied \cxlmem Latencies.}
\label{subsubsec:tpp-latency-sweep}
The variation in \cxlmem's access latency does not impact \TPP much.
We run Cache2 with 2:1 configuration where \cxlmem has different latency characteristics~(Figure~\ref{fig:tpp-latency-sweep}).
In all cases, due to {\TPP}'s better hot page identification and effective promotion, only a small portion (4--5\%) of hot pages are served from the \remote.
On the other hand, for default Linux, 22--25\% hot pages remain trapped in the \remote.
This increases the average memory access latency by $7\times$ for default Linux~(Figure \ref{fig:tpp-sweep-latency}).
This, eventually, impact the application-level performance, default Linux experiences $2.2-2.8\times$ higher throughput loss over \TPP ~(Figure \ref{fig:tpp-sweep-throughput}).

\subsection{Impact of \TPP Components}
\label{subsec:tpp-components}
In this section, we discuss the contribution of \TPP components.
As a case study, we use Cache1 with 1:4 configuration.

{\bf Allocation and Reclamation Decoupling.} 
Without this feature, reclamation on \local node triggers at a later phase.
With high memory pressure and delayed reclamation on \local node, the benefit of \TPP disappears as it fails to promote \remote pages.
Besides, newly allocated pages are often short-lived (less than a minute life-time) and de-allocated even before being selected as a promotion candidate.
Trapped {\remote} hot pages worsen the performance.

Without the decoupling feature, allocation maintains a steady rate that is controlled by the rate of reclamation (Figure~\ref{fig:tpp-cache1-wo-decoupling}). 
As a result, any burst in allocations puts pages to the {\remote}. 
When allocation and reclamation is decoupled, \TPP allows more pages on the \local node -- allocation rate to \local node increases by $1.6\times$ at the $95^{th}$ percentile.

As \local node is always under memory pressure, new allocations consume  freed up pages instantly and promotion fails as the target node becomes low on memory after serving allocation requests.
For this reason, without the decoupling feature, promotion almost halts most of the time (Figure~\ref{fig:tpp-cache1-w-decoupling}).
Trapped pages on the {\remote} generates 55\% of the memory traffic which leads to a 12\% throughput drop.
With the decoupling feature, promotion maintains a steady rate of 50KB/s on average.
During the surge in remote memory usage, the promotion  goes as high as 1.2MB/s in the $99^{th}$ percentile.
This 
helps \TPP to maintain the throughput of baseline.

\begin{figure}[!t]
	\includegraphics[width=0.9\columnwidth]{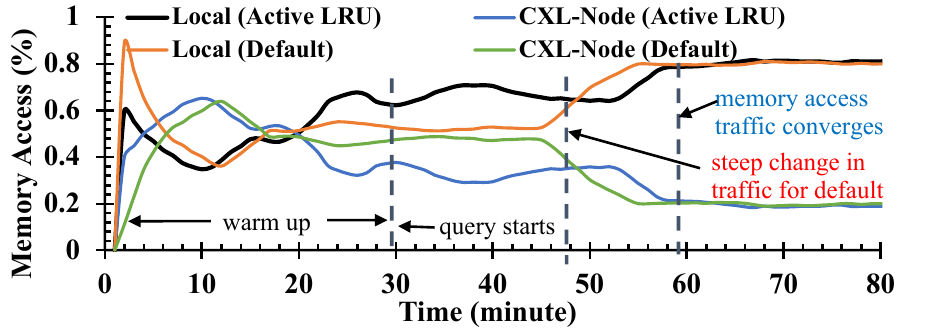}
	\caption{Restricting the promotion candidate based on their age reduces unnecessary promotion traffic.}
	\label{fig:eval-lru-promotion}
\end{figure}

\begin{table}[!t]
\caption{Page-type aware allocation helps applications.}
\resizebox{0.9\columnwidth}{!}{
\begin{tabular}{|c|c|cc|c|}
\hline
\multirow{2}{*}{Application} & \multirow{2}{*}{Configuration} & \multicolumn{2}{c|}{\% of Memory Access Traffic} & \multirow{2}{*}{\begin{tabular}[c]{@{}c@{}}Throughput\\ w.r.t Baseline\end{tabular}} \\ \cline{3-4}
 &  & \multicolumn{1}{c|}{\Local Node} & \remote &  \\ \hline
Web1 & 2:1 & \multicolumn{1}{c|}{97\%} & 3\% & 99.5\% \\ \hline
Cache1 & 1:4 & \multicolumn{1}{c|}{85\%} & 15\% & 99.8\% \\ \hline
Cache2 & 1:4 & \multicolumn{1}{c|}{72\%} & 28\% & 98.5\% \\ \hline
\end{tabular}
}
\label{table:tpp-remote-cache}
\end{table}

\begin{figure}[!t]
	\centering
	\subfloat[][\textbf{Web1 on 2:1 Configuration}]{
	\label{fig:tpp-autonuma-web}
		\includegraphics[width=0.48\columnwidth]{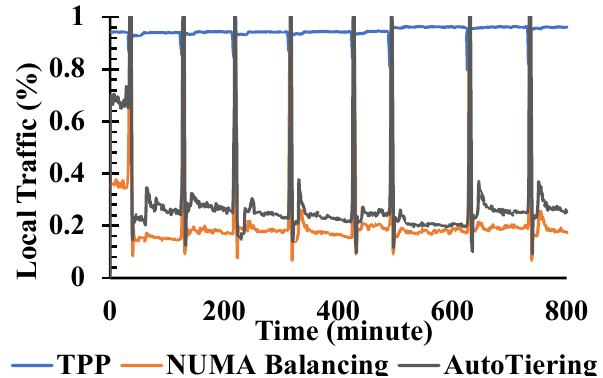}
	}
	\subfloat[][\textbf{Cache1 on 1:4 Configuration}]{
	\label{fig:tpp-autonuma-cache}
		\includegraphics[width=0.48\columnwidth]{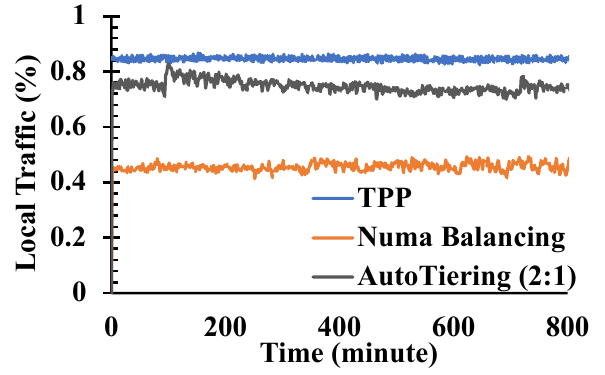}
	}
	\caption{\TPP outperforms existing page placement mechanism. Note that AutoTiering can't run on 1:4 configuration For Cache1, \TPP on 1:4 configuration performs better than AutoTiereing on 2:1.}
	\label{fig:tpp-autonuma}
\end{figure}

{\bf Active LRU-Based Hot Page Detection.} 
Considering active LRU pages as promotion candidates helps \TPP add hysteresis to page promotion.
This reduces the page promotion rate by $11\times$. 
The number of demoted pages that subsequently get promoted is also reduced by 50\%.
Although the demotion rate drops by 4\%, as we are not allowing unnecessary promotion traffics to waste local memory node, now there are more effective free pages in local node.
As a result, promotion success rate improves by 48\%.
Thus, reduced but successful promotions provide enough free spaces in local node for new allocations and improve the \local node's memory access by 4\%. 
Throughput also improves by 2.4\%.
The time requires to converge the traffic across memory tiers is almost similar -- to reach the peak traffic on \local node, \TPP with active LRU-based promotion takes extra five minutes (Figure~\ref{fig:eval-lru-promotion}).


{\bf Cache Allocation to Remote Node Policy.}
For Web and Cache, preferring the file cache allocation to {\remote} can provide all-local performances even with a small-sized \local node (Table~\ref{table:tpp-remote-cache}).
\TPP is efficient enough to keep most of the effective hot pages on the \local node.
Throughput drop over the baseline is only 0.2--2.5\%.

\subsection{Comparison Against Existing Solutions}
\label{subsec:tpp-counterparts}
\rev{We compare \TPP against Linux's default NUMA Balancing, AutoTiering, and TMO.}
We use Web1 and Cache1, two representative workloads of two different service domains, and evaluate them on  target production setup (2:1 configuration) and  memory-expansion setup (1:4 configuration), respectively.
We omit Data Warehouse as it does not show any significant performance drop even with default Linux (\S\ref{subsec:tpp-effectiveness}).

\begin{table}[!t]
\caption{\rev{TMO enhances \TPP's memory reclamation process and improves page migration by generating more free space.}}
\resizebox{0.9\columnwidth}{!}{
\begin{tabular}{|c|c|c|}
\hline
\rev{Web1 on 2:1 Configuration} & \rev{\TPP-only} & \rev{\TPP with TMO}  \\ \hline
\rev{Migration Failure Rate (pages/sec)} & \rev{20} & \rev{5}  \\ \hline
\rev{\remote's Memory Traffic (\%)} & \rev{3.1\%} & \rev{2.7\%}  \\ \hline
\end{tabular}
}
\label{table:tmo-help-tpp}
\end{table}

\begin{table}[!t]
\caption{\rev{\TPP improves TMO by effectively turning the swap action into a two-stage demote-then-swap process.}}
\resizebox{0.9\columnwidth}{!}{
\begin{tabular}{|c|c|c|}
\hline
\rev{Web1 on 2:1 Configuration} & \rev{TMO-only} & \rev{TMO with \TPP}  \\ \hline
\rev{Process Stall (Normalized to Threshold)} & \rev{70\%} & \rev{40\%}  \\ \hline
\rev{Memory Saving (\% of Total Capacity)} & \rev{13.5\%} & \rev{16.5\%} \\ \hline
\end{tabular}
}
\label{table:tpp-help-tmo}
\end{table}

\subsubsection{\TPP against NUMA Balancing and AutoTiering.}

\textbf{Web1.} NUMA Balancing helps when the reclaim mechanism can provide with enough free pages for promotion.
When the \local node is low on memory, NUMA Balancing stops promoting pages and performs even worse than default Linux because of its extra scanning and failed promotion tries.
Due to $42\times$ slower reclamation rate than \TPP, NUMA Balancing experiences
$11\times$ slower promotion rate.
Local node can serve only 20\% of the memory traffic (Figure~\ref{fig:tpp-autonuma-web}). 
As a result, throughput drops by 17.2\%.
Due to the unnecessary sampling, its system-wide CPU overhead is 2\% higher that \TPP.

AutoTiering has a faster reclamation mechanism -- it  migrates pages with low access frequencies to {\remote}.
However, with a tightly-coupled allocation-reclamation path, it maintains a fixed-size buffer to support promotion under pressure.
This reserved buffer eventually fills up during a surge in \remote page accesses.
At that point, AutoTiering also fails to promote pages and end up serving 70\% of the traffic from the {\remote}.
Throughput drops by 13\% from the baseline.

Note that \TPP experiences only a 0.5\% throughput drop.

{\bf Cache1.}
In 1:4 configuration, with more memory pressure on \local node, NUMA Balancing effectively stops promoting pages.
Only 46\% of the traffics are accessed from the \local node (Figure~\ref{fig:tpp-autonuma-cache}). 
Throughput drops by 10\%.

We can not setup AutoTiering with 1:4 configuration. 
It frequently crashes right after the warm up phase, when query fires.
We run Cache1 with AutoTiering on 2:1 configuration.
\TPP out performs AutoTiering even with a 46\% smaller \local node -- \TPP can serve 10\% more traffic from \local node and provides 7\% better throughput over AutoTiering.

\subsubsection{Comparison between \TPP and TMO}
\label{subsubsec:tpp-tmo} 
\textbf{\TPP vs. TMO.} \rev{TMO monitors application stalls during  execution time because of insufficient system resources (CPU, memory, and IO).
Based on the pressure stall information (PSI), it decides on the amount of memory that needs to be offloaded to the swap space. 
As mentioned in \S\ref{sec:tpp-principal}, using TMO for \cxlmem's (z)swap-based abstraction is beyond our design goal.
For the sake of argument, if we configure \cxlmem as a swap-space for TMO, it will be only populated during reclamation. 
New page allocation can never happen there. 
Besides, without any fast promotion mechanism, aggressive reclamation can hurt application's performance; especially, when reclaimed pages are re-accessed through costly swap-ins. 
As a result, TMO throttles and can not populate most of the \cxlmem capacity. 
For Web1, Cache1, and Data Warehouse in 2:1 configuration, TMO can only consume 45\%, 61\%, and 7\% of the \cxlmem capacity, respectively.
On the other hand, \TPP can use \cxlmem for both allocation and reclamation purposes. 
For same applications, \TPP's \cxlmem usage is 83\%, 92\%, and 87\%, respectively.}

\textbf{\TPP with TMO.} 
\rev{We run TMO with \TPP and observe they are orthogonal and augment each other’s behavior for Web1 on 2:1 configuration.}
\rev{\TPP keeps most hot pages in \local node; 
it gets slightly better when TMO is enabled (Table \ref{table:tmo-help-tpp}).}
\rev{TMO creates more free memory space in the system by swapping out cold pages both from local and \cxlmem nodes.}
\rev{The presence of some memory headroom in the system makes it easier for \TPP to move pages around and leads to fewer page migration failures.}
\rev{As \TPP-driven migration fails less frequently, page placement is more optimized, resulting in even fewer accesses to the \remote.}

\rev{TMO is still able to save memory without noticeable performance overhead, as shown in Table \ref{table:tpp-help-tmo}.}
\rev{This is because \TPP makes (z)swap a two-stage process -- TMO-driven reclamation in local node will first demote victim pages to \cxlmem before getting (z)swapped out eventually.}
\rev{This improves victim page selection process -- semi-hot pages now get a second chance for staying in local memory when drifting down \remote's LRU list.
As a result, \TPP reduces the amount of process stall in TMO originated from major page faults (\ie, memory and IO pressure in \cite{tmo}) by 30\%.
As TMO throttles itself based on the process stall metric, this improves memory saving by 3\% (2GB in absolute terms).}
\section{Discussion and Future Research}
\label{sec:discussion}

\TPP makes us production-ready to onboard our first generation of CXL-enabled tiered-memory system.
We, however, foresee research opportunities with technology evolution.

{\bf Tiered Memory for Multi-tenant Clouds.}
In a typical cloud, when multiple tenants co-exist on a single host machine, \TPP  can effectively enable them to competitively share different memory tiers. 
When local memory size dominates the total memory capacity of the system, this may not cause much problem.
However, if applications with different priorities have different QoS requirements, \TPP may provide sub-optimal performance. 
Integrating a well-designed QoS-aware memory management mechanism over \TPP can address this problem.

{\bf Allocation Policy for Memory Bandwidth Expansion.}
For memory bandwidth-bound applications, CPU to DRAM memory bandwidth often becomes the bottleneck.
CXL's additional memory bandwidth can help 
by spreading memory across the top-tier and remote node. 
Instead of only placing cold pages into \cxlmem, which draw very low bandwidth consumption, the optimal solution should place the right amount of bandwidth-heavy, latency-insensitive pages to \cxlmem. 
The methodology to identify the ideal fraction of such working sets may even require hardware support.
%
We want to explore transparent memory management for memory bandwidth-expansion use case in our future work. 

{\bf Hardware Support for Effective Page Placement.}
Hardware features can further enhance performance of \TPP. 
A memory-side cache and its associated prefetcher on the CXL ASIC might help reduce the effective latency of \cxlmem.
\rev{Hardware support for data movement between memory tiers can help reduce page migration overheads. While in our environment we do not see a high migration overheads, others may chose to put provision systems more aggressively with very small amount of local memory and high amount of \cxlmem. For our use cases, in steady state, the migration bandwidth is 4--16 MB/s (1--4K pages/second) which is far lower than CXL link bandwidth and also unlikely to cause any meaningful CPU overhead due to page movement.}

\section{Related Work}
{\bf Tiered Memory System.}
With the emergence of low-latency non-DDR technologies, heterogeneous memory systems are becoming popular.
There have been significant efforts in using NVM to extend main memory~\cite{data-tiering, fb-dcpmm, baidu-optane, thermostat, die-stacked, hemem, heteroos, memverge, monet}.
CXL enables an intermediate memory tier with DRAM-like low-latency  in the hierarchy and brings a paradigm shift in flexible and performant server design.
Industry leaders are embracing CXL-enabled tiered memory system in their next-generation datacenters~\cite{intel-cxl, amd-cxl, micron-cxl, samsung-cxl, next-plarform-cxl, lenovo-cxl, single-socket-cxl}. 

{\bf Page Placement for Tiered Memory.}
Prior work explored 
hardware-assisted~\cite{hw-pcm-placement, hetero-mem, hybrid-mem-hotos} and application-guided~\cite{program-hybrid-mem, memkind, data-tiering, samsung-cxl} page placement for tiered memory systems, which may not often scale to datacenter use cases as they require hardware support or application redesign from the ground up. 

Application-transparent page placement approaches often  profile an application's physical~\cite{kleio, heteroos, thermostat, googledisagg, mrc-memory-management, damon-reclaim} or virtual address-space~\cite{utility-based-memory, chopt, nvm-runtime-manage, hemem} to detect page temperature. This causes high performance-overhead because of frequent invocation of TLB invalidations or interrupts. 
We find existing in-kernel LRU-based page temperature detection is good enough for \cxlmem.
Prior study also explored machine learning directed decisions~\cite{kleio, googledisagg}, user-space APIs~\cite{hemem,utility-based-memory}, and swapping~\cite{thermostat, googledisagg} to move pages across the hierarchy, which are either resource or latency intensive. 

In-memory swapping~\cite{zswap,frontswap,zram,gcpswap} can be used to swap-out cold pages to \remote.
In such cases, \remote access requires page-fault and swapped-out pages are immediately brought back to main memory when accessed. 
This makes in-memory swapping ineffective for workloads that access pages at varied frequencies. When \cxlmem is a part of the main memory, less frequently accessed pages can be on \remote without any page-fault overhead upon access.

Solutions considering NVM to be the slow memory tier~\cite{nimble, autotiering, top-tier-management, demotion} are conceptually close to our work.
Nimble~\cite{nimble} is optimized for huge page migrations.
During migration, 
it employs page exchange between memory tiers.
This worsens the performance as a demotion needs to wait for a promotion in the critical path.
Similar to \TPP, AutoTiereing~\cite{autotiering} and work from Huang et al.~\cite{top-tier-management} use background migration for demotion and optimized NUMA balancing~\cite{autonuma} for promotion.
However, their timer-based hot page detection causes computation overhead and is often inefficient, especially when pages are infrequently accessed.
Besides, none of them consider decoupling allocation and reclamation paths.
Our evaluation 
shows, this is critical for memory-bound applications to maintain their performance under memory pressure. 
 

{\bf Disaggregated Memory.} 
Memory disaggregation exposes capacity available in remote hosts as a pool of memory shared among many machines. 
Most recent memory disaggregation efforts~\cite{remote-regions,memtrade, infiniswap, leap, hydra, kona, LegoOS, AIFM, semeru, farmemory-throughput, alibaba-disagg} are specifically designed for RDMA over InfiniBand or Ethernet networks where latency characteristics are orders-of-magnitude higher than \cxlmem.
Memory managements of these systems are orthogonal to \TPP -- one can use both CXL- and network-enabled memory tiers and apply \TPP and memory disaggregation solutions to manage memory on the respective tiers.



\section{Conclusion}
We analyze datacenter applications' memory usage behavior using \pebstool, a lightweight and robust user-space working set characterization tool, to find the scope of CXL-enabled tiered-memory system.
To realize such a system, we design \TPP, an OS-level transparent page placement mechanism that works without any prior knowledge on applications' memory access behavior.
We evaluate \TPP using diverse production workloads and find \TPP improves application's performance on default Linux by 18\%. 
\TPP also outperforms NUMA Balancing and AutoTiering, two state-of-the-art tiered-memory management mechanisms, by 5--17\%.
\section*{Acknowledgments}
\label{sec:acknowledgments}
\rev{
We thank the anonymous reviewers for insightful feedback 
that helped improve the paper. 
Hasan Al Maruf and Mosharaf Chowdhury were partly supported by National Science Foundation grants (CNS-1845853, CNS-2104243) and gifts from VMware and Meta.
}

\label{lastpage}
\bibliographystyle{abbrv}
\bibliography{cxl-memory}
\end{document}